# Electron ptychography reveals a ferroelectricity dominated by anion displacements

Harikrishnan KP,[1] Ruijuan Xu,[2,3,4] Kinnary Patel,[5] Kevin J. Crust,[2,6] Aarushi Khandelwal,[2,3] Chenyu Zhang,[1] Sergey Prosandeev,[5] Hua Zhou,[7] Yu-Tsun Shao,[1,8] Laurent Bellaiche,[5,9] Harold Y. Hwang,[2,3] David A. Muller[1,10,*]

[1.] School of Applied and Engineering Physics, Cornell University, Ithaca, NY, USA.
[2.] Stanford Institute for Materials and Energy Sciences, SLAC National Accelerator Laboratory, Menlo Park, CA, USA.
[3.] Department of Applied Physics, Stanford University, Stanford, CA, USA.
[4.] Department of Materials Science and Engineering, North Carolina State University, Raleigh, NC, USA.
[5.] Smart Ferroic Materials Center, Physics Department and Institute for Nanoscience and Engineering, University of Arkansas, Fayetteville, Arkansas 72701, USA.
[6.] Department of Physics, Stanford University, Stanford, CA, USA.
[7.] X-ray Science Division, Advanced Photon Source, Argonne National Laboratory, Lemont, IL, USA.
[8.] Mork Family Department of Chemical Engineering and Materials Science, University of Southern California, Los Angeles, CA, USA.
[9.] Department of Materials Science and Engineering, Tel Aviv University, Ramat Aviv, Tel Aviv 6997801, Israel.
[10.] Kavli Institute at Cornell for Nanoscale Science, Ithaca, NY, USA.
*Corresponding Author: E-mail: david.a.muller@cornell.edu

## Abstract

Sodium niobate, a lead-free ferroic material, hosts delicately-balanced, competing order parameters, including ferroelectric states that can be stabilized by epitaxial strain. Here, we show that the resulting macroscopic ferroelectricity exhibits an unconventional microscopic structure using multislice electron ptychography. This technique overcomes multiple scattering artifacts limiting conventional electron microscopy, enabling both lateral spatial resolution beyond the diffraction limit and recovery of three-dimensional structural information. These imaging capabilities allow us to separate the ferroelectric interior of the sample from the relaxed surface structure and identify the soft phonon mode and related structural distortions with picometer precision. Unlike conventional ferroelectric perovskites, we find that the polar distortion in this material involves minimal distortions of the cation sublattices and is instead dominated by anion displacements relative to the niobium sublattice. We establish limits on film thickness for interfacial octahedral rotation engineering and directly visualize a random octahedral rotation pattern, arising from the flat dispersion of the associated phonon mode.

## Main

Sodium niobate ($NaNbO_3$) has received significant research attention due to its potential as a lead-free candidate material for energy storage applications[1,2]. $NaNbO_3$ is known for its complex structures, accompanied by a variety of tilt and rotation patterns of the oxygen sites, resulting in a rich phase diagram exhibiting multiple competing ground states and temperature-driven phase transitions[3–9] (Supplementary Text 1). While $NaNbO_3$ exhibits an antiferroelectric state in bulk at room temperature, a macroscopic room-temperature ferroelectric state can also be stabilized in $NaNbO_3$ thin films through epitaxial strain[10–20]. Elucidating the structural nature of this strain-stabilized ferroelectric phase in $NaNbO_3$ requires an accurate imaging of both cation and anion sublattices for dipolar mapping.

More generally, understanding the structural distortions at oxygen sites is crucial not only in the study of ferroelectricity but also for other emergent properties like charge and orbital ordering, and associated metal-insulator transitions in complex oxides[21–23]. Therefore, an accurate characterization of oxygen atoms is critical in understanding the structure-property relationship in oxide materials. However, this task is often complicated by the weak scattering power of the light oxygen atoms, making both X-ray and electron scattering techniques less sensitive to oxygen in the presence of heavier cations. In the study of ferroelectrics, cation displacements alone are often used as a proxy to estimate polarization, based on the assumption that the polarization correlates with the displacement of one cation sublattice with respect to another. This approach has proven to be qualitatively effective for the now well-studied displacive ferroelectrics[24–26] where the potentially large displacements of the oxygen atoms with respect to the cations are often inferred indirectly from the cation-dominated measurements in scattering experiments. However, measuring only cation-cation displacements, which is common in electron microscopy, overlooks the essence of a dipole moment, which requires an offset between the centroid of cations and anions. Moreover, such a qualitative approach does not easily generalize and overlooks other mechanisms, including those dominated by anion displacements relative to a particular cation sublattice as we show here in the case of tensile-strained $NaNbO_3$ grown epitaxially on $DyScO_3$. In the context of $NaNbO_3$, when we say anion displacements, we mean the net displacement of the centroid of the oxygen atoms with respect to the centroid of the Nb atoms for each pseudo-cubic unit cell, which forms one of the dipoles in the material.

In this study, we use multislice electron ptychography (MEP)[27,28], a new imaging modality in scanning transmission electron microscopy (STEM) that solves the inverse problem of electron scattering to retrieve the sample potential in three dimensions with both lateral and depth resolution beyond the diffraction limit. By accounting for multiple scattering and channeling effects that limit conventional STEM techniques, MEP enables us to reliably map out the light oxygen atoms in $NaNbO_3$ with picometer precision. This ability to accurately characterize the oxygen positions proved key to identifying an unconventional polar distortion dominated by anion displacements relative to the Nb sublattice, critical for the microscopic characterization of ferroelectricity in this material.

While not detectable with conventional STEM imaging modes, using MEP, we find that the ferroelectric phase of NaNbO$_3$ stabilized by epitaxial strain from a DyScO$_3$ substrate exhibits large oxygen octahedral rotations (OOR) and anti-polar cation displacements, both of which have previously been found to suppress the ferroelectric mode[29,30]. However, biaxial tensile strain from epitaxy induces an in-plane polarization in NaNbO$_3$[12,17], resulting in the simultaneous stabilization of these traditionally incompatible modes: OOR and ferroelectricity or anti-polar cation displacements and ferroelectricity, with the polarization arising from an unconventional microscopic distortion dominated by anion displacements relative to the Nb sublattice and with minimal cation-cation distortions. This finding is notably different from prototypical displacive ferroelectrics, where the inversion symmetry breaking is measured from and hence traditionally associated with large cation-cation displacements. Conventional cation-dominated structural characterization would mistake NaNbO$_3$ as nearly centrosymmetric, as the net cation displacement not only falls below the detection threshold of most atomic-scale scattering techniques but is also convolved with centrosymmetry-preserving anti-polar displacements of a larger magnitude.

The MEP measurements also explain why conventional STEM imaging fails to detect the ferroelectric state in thin sections of NaNbO$_3$, as the sections have a relaxed surface structure where the OOR is suppressed. Conventional STEM methods are most sensitive to the top surface of the sample[31] and hence detect the relaxed surface structure, whereas the three-dimensional (3D) structural resolution achieved through MEP[32–34] enables us to detect and disentangle the relaxed surface structure from the ferroelectric interior of the sample. This depth-sectioning capability also enables us to detect buried nano-regions in the sample where the OOR switches direction while the polarization does not, showing that the OOR and polarization order parameters are not directly coupled at relatively low orders (higher order coupling terms cannot be entirely ruled out).

In addition to free surfaces, we would also expect relaxation of OOR at interfaces with materials that have different octahedral tilt patterns, as is the case for the $NaNbO_3$/$DyScO_3$ interface. Again, using ptychographic imaging of interfaces, we map the oxygen octahedral coupling between the substrate and film, identifying the governing length scale for this coupling and setting thickness limits for octahedral tilt engineering, where octahedral tilts of the substrate are templated into the film. Using MEP, we also record visual evidence for random OOR patterns in $NaNbO_3$ which arise from an unusually flat phonon branch between points in the Brillouin zone that drive in-phase and anti-phase tilting of the octahedra[35,36].

**Comparison of ptychography with conventional STEM methods**

Figure 1(a) illustrates the experimental setup for MEP. In this technique, as the probe is scanned across the sample, the full diffraction pattern is collected at each scan point using a pixelated detector, leading to a four-dimensional dataset (two diffraction space and two scan/real space coordinates). Phase information encoded in the diffraction patterns from the overlap of scattered beams is retrieved using an iterative algorithm, enabling the reconstruction of the full scattering potential[28]. The forward model for the propagation of the electron beam uses the multislice method to solve Schrödinger's equation[27], where the sample is divided into different slices along the beam direction to account for multiple scattering of the electron beam, and changes in probe shape during propagation. The MEP reconstruction (solution to the inverse problem) recovers the potential for each of these slices individually, hence providing information about the sample structure along the beam direction.

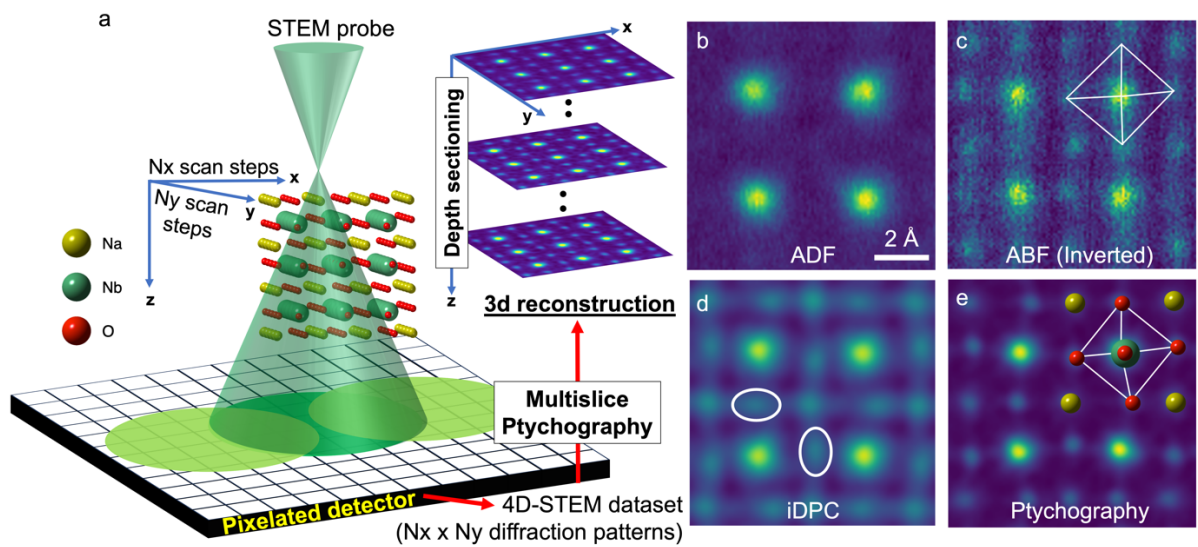

**Fig. 1| Schematic of multislice electron ptychography and comparison with conventional STEM techniques.** *a, Schematic of the experimental setup for MEP and reconstruction workflow. Comparison of (b) ADF, (c) inverted ABF, (d) iDPC and (e) projected MEP images for a 2x2 pseudo-cubic unit cell region of $NaNbO_3$. The ADF image is sensitive only to the heavy Nb atoms and fails to capture the light Na and O atoms. The ABF image is blurred and shows a substantially damped OOR compared to that in the MEP image. The iDPC image shows blurring of the oxygen columns in the direction corresponding to the rotation of the octahedra as marked by the white ovals. The MEP image captures the structure of the sample at a resolution and precision that conventional techniques cannot match. Even subtle details like the variations in the different Nb-O bond lengths and angles are directly detectable in the image.*

Figure 1(b-e) shows a direct comparison of conventional STEM imaging techniques such as annular dark field (ADF), annular bright field (ABF), and integrated differential phase contrast (iDPC) with the MEP image of a 2 × 2 pseudo-cubic unit-cell region of the $NaNbO_3$ thin film (see Extended Data Fig. 1 for larger field of view). With no information about the positions of the lighter sodium or oxygen atoms, ADF imaging has limited utility for quantitative characterization in this system. With the typical precision of ADF imaging (~5 pm)[37], the niobium atoms appear to form a cubic lattice, misleadingly suggesting an absence of polar order. Even with enough precision to measure deviations from cubic symmetry in the niobium sublattice, it would only reveal an antiferroelectric ordering of niobium atoms. In contrast, the ABF and iDPC images pick out the lighter sodium and oxygen atoms too, albeit with inferior resolution and contrast compared to the MEP reconstruction. A comparison of these images with the ptychographic reconstruction also reveals discrepancies in the measured positions of the oxygen atoms, with an incorrect, reduced magnitude of the OOR in the ABF image and elongated oxygen columns (marked with white ovals) in the iDPC image, making any quantification of the atomic positions problematic. ABF images taken at different regions of the sample can show large structural variations primarily due to changes in the sample thickness, making interpretation of individual images unreliable – examples of such variations are shown in Extended Data Fig. 2. Previous studies have also demonstrated ADF[37–39], ABF[40,41] and iDPC[42,43] images to produce significant systematic errors in atomic positions even from small variations in diffraction conditions that are unavoidable in practical experimental conditions.

In comparison to the imaging modes discussed above, the enhanced resolution in the ptychographic image is visually apparent and facilitates a very precise quantification of the atomic positions, critical in understanding the different structural distortions. The variations in the Nb-O bond lengths and angles are directly interpretable from the image without the need for complicated modeling or fitting. We also note that the apparent width of the atoms is proportional to their atomic number and not measurably sensitive to the ionic radius. In the next section, we will discuss how the structural resolution along the projection direction enabled by

MEP addresses the misplacement of atoms seen in ABF imaging and the elongation of oxygen columns in the iDPC image.

## Tracking structural changes in three dimensions

There are three different surfaces in the thin TEM lamella we made – the original top surface, and two new surfaces from preparing a cross-sectional lift-out. Here, we will discuss the newly introduced surfaces of the lamella in the cross-sectional lift-out (which become the entrance and exit surfaces of the electron beam) rather than the top surface of the film. The depth-sectioning capability of MEP reveals a surface relaxation in the sample and disentangles the presence of multiple phases along the depth direction as illustrated in Fig. 2. The depth profile of a pseudo-cubic $(001)_{pc}$ Na-O plane along the $[100]_{pc}$ direction in a cross-sectional $NaNbO_3$ sample, shown in Fig. 2(b), reveals an offset in the position of the oxygen atoms at the entrance and exit surfaces of the sample with respect to their positions in the interior. The structure of the interior and the surface of the sample (obtained by summing up the slices within the labelled yellow boxes) is illustrated in Fig. 2(a) and (c), showing a reduced magnitude of the OOR at the lamella surface, arising from surface relaxations (quantitative results shown in Supplementary Fig. 1). The Na-O plane from which the depth profile is extracted is also labelled with a dashed yellow box in Fig. 2(a) and (c). By removing the slices that correspond to the top and bottom relaxed lamella surfaces, one can obtain measurements reflecting the sample's interior bulk structure. Separating out the surface effects is particularly important if the measured properties are to be compared with other bulk measurements. Moreover, the study of the reconstructed surface structure itself can be of interest for catalytic applications[44] or the exploration of emergent quantum phenomena[45].

A more pronounced surface relaxation was observed in sodium niobate films grown on strontium titanate, shown in Extended Data Fig. 3. Using simulations shown in Extended Data Fig. 4, we also show how such surface relaxations could explain the source of artifacts in the ABF and iDPC images shown in Fig. 1. Both ABF and iDPC images have the best contrast when the beam is focused near the top surface of the sample[31,46] and therefore convey information about the relaxed surface structure. Under conditions where conventional STEM modalities provide the best contrast, they fail to reveal the correct interior structure of the sample.

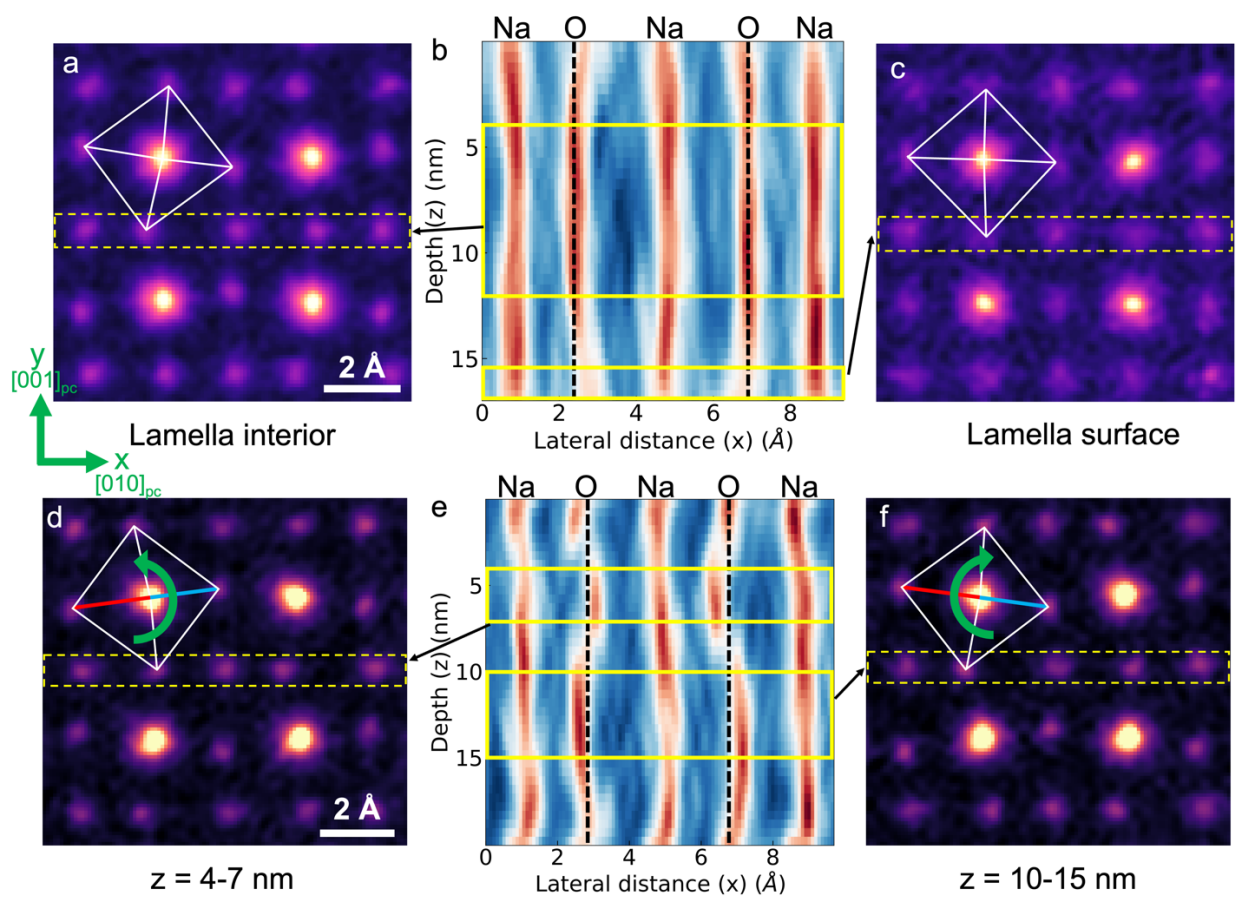

**Fig. 2 | Depth sectioning reveals surface relaxation and nanoscale phases in projection.**
*Ptychographic reconstruction zoomed in to a 2x2 pseudo-cubic unit cell region of NaNbO₃ showing the (a) interior and (c) surface structure of the sample. The magnitude of the OOR is suppressed at the lamella surface compared to that in the interior of the sample. b, Depth profile along the Na-O plane marked with a yellow dotted box in (a, c) shows the deviation in the*

*position of the oxygen atoms at both lamella surfaces in comparison to the interior. The slices summed to get the images shown in (a, c) are also labelled in (b).* ***d-f,*** *Ptychographic image of NaNbO$_3$ spatially separated from the region shown in (a-c).* ***e,*** *Depth sectioning along the Na-O plane labelled with a yellow dotted box in (d, f) shows multiple nanoscale phases present in the projection direction arising from large shifts in the position of the oxygen atoms. These phases differ in the sense of OOR as shown in **(d, f)** obtained by summing the slices corresponding to the two different phases. The blue and red solid lines label the short and long in-plane Nb-O bond lengths respectively and show that the polarization direction does not switch when the octahedral rotation switches direction, indicating that the order parameters associated with the two effects are not directly coupled.*

In addition to the relaxation at the lamella surface, samples often exhibit multiple OOR phases along the depth direction, which is illustrated in Fig. 2(d-f). Fig. 2(e) shows the depth profile along a pseudo-cubic (001) Na-O plane and shows large shifts in the position of the oxygen atoms in the interior of the sample, because of a reversal in the direction of OOR. At approximately 3 nm depth into the sample, the rotation for each octahedron switches sign for approximately 6 nm before returning to the original configuration. Fig. 2(d) and (f) detail the structure of these nano-regions, where the rotation direction of oxygen octahedra alternates, located at different depths in the sample. Blue and red solid lines are used to denote the short and long in-plane Nb-O bond lengths, respectively, and indicate that the polarization direction remains unchanged across both nano-regions despite the reversal in the OOR. The unchanged polarization amidst a switch in the sign of the octahedral rotation suggests that the associated order parameters are not directly coupled at relatively low orders (potential higher-order interactions cannot be dismissed). However, the presence of nano-regions with alternating octahedral tilts also indicates a low-energy cost associated with displacements of the oxygen atoms, an observation that will become relevant in the discussion of polarization. This analysis demonstrates how depth sectioning with MEP allows the segmentation and characterization of nanoscale phases within the sample depth, which would be overlooked with the conventional imaging modes.

## Polarization from anion displacements

As motivated in the above discussion, it is the accurate characterization of the oxygen positions and disentanglement from the relaxed surface structure, that enables a robust characterization of the NaNbO$_3$/DyScO$_3$ heterostructure. The 58 nm-thick films have high crystalline quality and possess coherent tensile strain to the substrate as can be seen in Extended Data Fig. 5. Figure 3 (a-e) shows schematics of the primary distortions in the system in terms of irreducible representations of the parent cubic perovskite structure ($Pm\bar{3}m$), projected along a cubic axis in Fig. 3(a). Pure rotations of the oxygen octahedra along a cubic axis result in a doubling of the unit cell size along the two orthogonal axes due to the three-dimensional connectivity of the octahedra, driven by zone-boundary modes. If the direction of rotation stays constant along the axis (in-phase rotation), the relevant mode is $M_3^+$, while if the direction switches between neighboring unit cells along the axis (anti-phase rotation), the relevant mode is $R_4^+$ (Supplementary Fig. 2). Such octahedral rotations preserve centrosymmetry as shown in Fig. 3(b) for the case of an in-phase octahedral rotation along the projection direction. The centrosymmetry in the system is broken by a ferroelectric displacive mode of the oxygen atoms that transforms as the zone-center $\Gamma_4^-$ mode and creates a distorted octahedra as shown in Fig. 3(c). These distortions can be directly visualized in the experimental ptychography reconstruction shown in Fig. 3(d) where an overlay of purely tilted octahedra does not match the experimentally measured oxygen atom positions. In addition to the OOR, the oxygen atoms show an almost rigid in-plane shift with respect to the cations that generates the dipole moment as depicted in Fig. 3(c). Although such an octahedral distortion can be equivalently explained by a displacement of the B-site cation, the niobium atom remains almost centered with respect to the surrounding sodium atoms. Essentially, the cation sublattices maintain near-perfect centrosymmetry, and the inversion symmetry is broken largely by anion displacements with respect to the Nb sublattice. The earlier observation in Fig. 2 of the oxygen atom displacements dominating the surface relaxation and the presence of nano-regions with alternating octahedral rotations also suggests the low-energy cost involved in displacing the oxygen atoms. In addition to the distortion of the oxygen octahedra, the sodium and niobium cations also show small anti-polar displacements (transforming like the zone-boundary $M_5^-$ mode) along the in-plane and out-of-plane directions and with magnitudes of 6.2 pm and 2 pm respectively in projection, as shown in Fig. 3(e). The detailed description of the different lattice distortions and associated satellite peaks in the Fourier domain when imaged along both pseudo-cubic axes can be found in Extended Data Fig. 6 and Supplementary Fig. 3.

Despite having a tolerance factor (=0.974) close to 1, $NaNbO_3$ still exhibits large octahedral tilts with a magnitude of around 7.3 degrees about both in-plane pseudo-cubic axes (see Extended Data Fig. 7 for histograms of the measured octahedral tilts). Prior theoretical work has established that octahedral rotations and anti-polar A-site displacements that generally stabilize the A-site coordination suppress the polar mode[30]. Despite the presence of anti-polar displacements of both cation sublattices and large octahedral rotations, the tensile-strained $NaNbO_3$ develops a robust in-plane polarization primarily arising from an anion displacement dominated Γ point mode of the parent cubic perovskite.

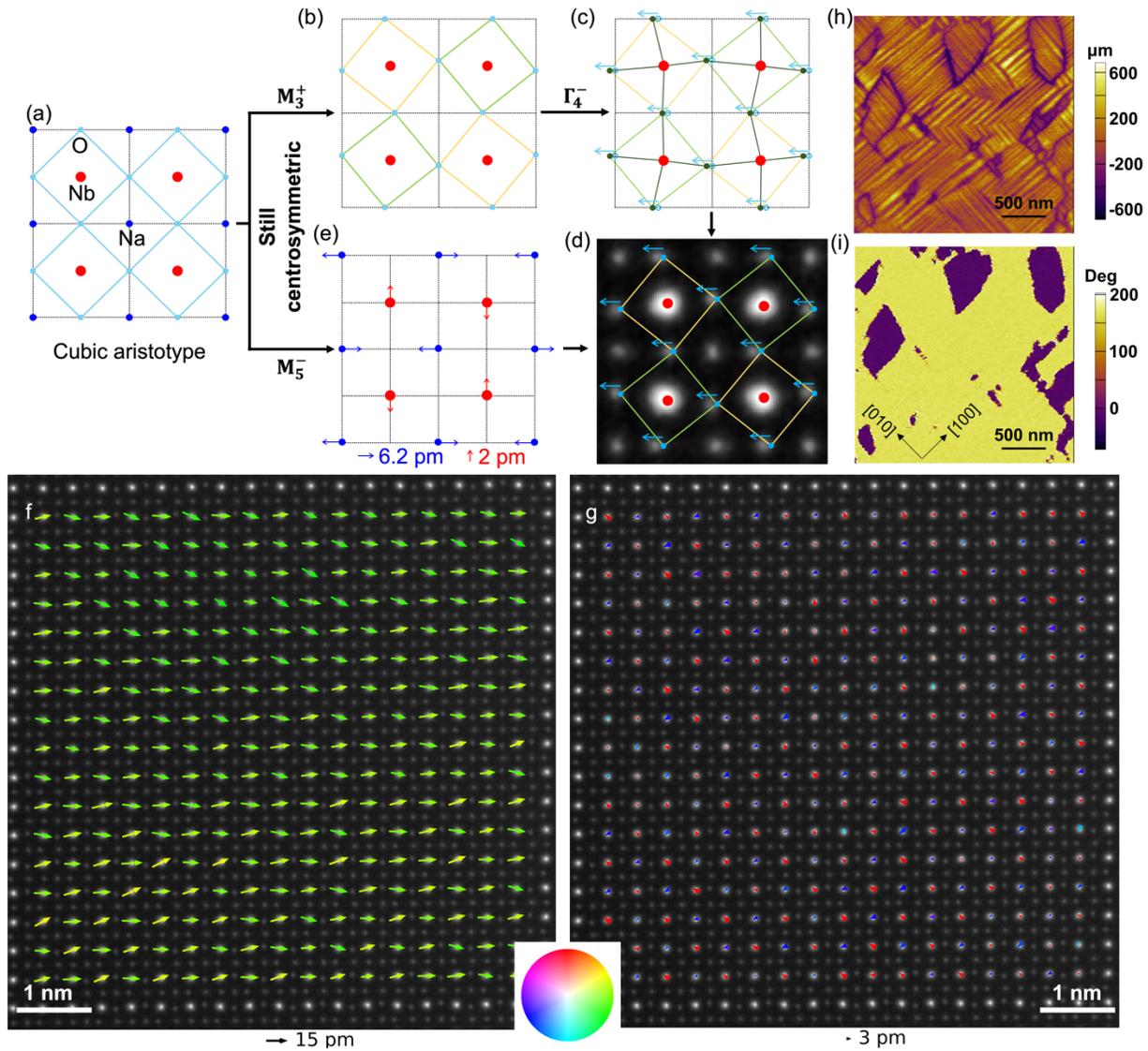

**Fig. 3 | Visualizing structural distortions that result in the microscopic dipole moment and characterization of the macroscopic polarization.** *a-c, Schematic summarizing the different structural distortions in NaNbO$_3$ starting from a 2x2 unit cell undistorted perovskite (a). The structure after imposing octahedral rotations (b) and additional displacements of the oxygen atoms (c) is compared to an experimental ptychographic image (d) with an overlay of NbO$_6$ octahedra with a pure rotation. The oxygen positions in the presence/absence of the octahedral distortion are shown using green/blue dots, with arrows used to show the direction of oxygen displacement. The tail of the arrow lines up with the blue dots and the length is exaggerated for better visualization. The oxygen displacements seen as an offset in the experimentally measured oxygen positions with respect to the undistorted octahedra in (d) results in a charge separation creating the polarization. e, Schematic showing the anti-polar displacements exhibited by the cations – Na atoms show alternating left/right displacements while Nb atoms show alternating up/down displacements. f, Vector map between the centroid of the oxygen atoms and the Nb atom for each pseudo-cubic unit cell showing predominantly in-plane polarization. g, Map of the*

*displacement of the Nb atom with respect to the centroid of the Na atoms for each pseudo-cubic unit cell that fails to capture the details of the polarization.* **h-i,** *Lateral PFM amplitude (h) and phase (i) images of a 2.5 x 2.5 um region of the sample showing the well-ordered domain structure.*

Figure 3(f) shows a map of the displacement between the centroid of the negatively charged oxygen atoms and the positively charged Nb atom for each pseudo-cubic unit cell. This displacement map indicates a predominantly in-plane polarization, with the small up-down canting of the arrows related to the anti-polar displacements of the niobium atoms. The map of the other dipole formed from the displacement of the centroid of the sodium atoms from the centroid of the oxygen atoms is shown in Extended Data Fig. 8, along with displacement maps measured along the other in-plane pseudo-cubic direction. Together, these displacement maps indicate that the polarization points predominantly in-plane roughly along the $[110]_{pc}$ direction with a small out-of-plane component, consistent with our piezo-force microscope (PFM) measurements. The lateral PFM imaging (amplitude and phase shown in Fig. 3 (h, i) respectively) shows a stronger piezoelectric response with a more well-defined domain structure in comparison to the vertical PFM results (see Extended Data Fig. 9 for further details about the PFM characterization and different domain variants). The formation of such an in-plane polydomain structure would reduce the depolarization energy caused by the in-plane polarization. Given that there are three free surfaces (two lamella surfaces and the top surface), the surface charge from the polarization is likely screened by adsorbed molecules or oxygen and sodium vacancies. The average domain size from PFM characterization is found to be around 50 nm, which is in rough agreement with the results extracted from the 3D reciprocal space map shown in Extended Data Fig 5. We also detect the presence of the overall net out-of-plane polarization by the breaking of Friedel symmetry in electron diffraction patterns (Supplementary Fig. 4).

From the MEP data in Fig. 3 and Extended Data Fig. 8, the measured lengths of the structural dipoles are 12.5±1.1 (11.1±1.7) pm for Nb-O and 15.3±1.5 (13.9±2.1) pm for Na-O when projected down the $[1\bar{1}0]_O$ ($[001]_o$) zone axis of the DyScO$_3$ substrate, which would result in a net in-plane polarization of 27.8±3.3 μC/cm$^2$ rotated 3.2° away from the $[110]_{pc}$ direction if we naively assume formal charges (+1 on Na, +5 on Nb and -2 on O) on the atoms. In comparison to this toy model, density functional theory (DFT) calculations (on candidate structures that will be introduced in the next section) produce a total polarization around 45 μC/cm$^2$ along the $[110]_{pc}$ direction with large Born effective charges on the Nb (> 9e$^-$) and O (> 7e$^-$) atoms as expected for ferroelectrics (see Supplementary Text 3 for more details). The measured out-of-plane polarization is observed to be spatially inhomogeneous and is an order of magnitude lower than the in-plane polarization.

We emphasize that it is the robust imaging of the oxygen atoms with MEP that enabled the application of the physical definition of the dipole moment to extract the displacement map shown in Fig. 3(f). If instead, the maps were calculated solely from the displacement of the niobium cations with respect to the centroid of the sodium cations as shown in Fig. 3(g), the resulting displacements would be about 5 times smaller and not as well ordered as in Fig. 3(f) while also carrying an overall opposite sign. Hence, the current materials system represents an example where the often-used assumption in electron microscopy of the polarization being proportional to the displacement of one cation sublattice with respect to the other breaks down, and where tracking the displacements of oxygen atoms with respect to the Nb sublattice is key in accurately characterizing the polarization.

**New monoclinic phases**

In Fig. 4(a, c), we show a schematic of anti-phase and in-phase OOR and illustrate how they can be differentiated in projection by the presence of double or single oxygen columns respectively. We demonstrate this capability with MEP images of the DyScO$_3$ substrate along the $[1\bar{1}0]_O$ and $[001]_O$ zone axes in Fig. 4(b, d). The observation of double and single oxygen columns in Fig. 4 (b, d) respectively is consistent with the known OOR pattern along these zone axes – anti-phase for $[1\bar{1}0]_O$ and in-phase for $[001]_O$. For the phase of NaNbO$_3$ discussed above, the MEP images along both in-plane pseudo-cubic directions display a single oxygen column as shown in Fig. 4(e), enabling us to infer an in-phase OOR pattern. Using DFT calculations to search for phases with in-phase tilting, we reproduced monoclinic structures of NaNbO$_3$ that match the experimentally determined structure – $Pc$ phase with tilt configuration of $a^+b^+c^-$ and $Cm$ phase with tilt configuration of $a^+b^+c^+$, where the tilt configurations are denoted with the Glazer notation[47]. The above octahedral tilt patterns are relatively rare among perovskites and the symmetries of these ferroelectric phases are distinct from the low temperature $R3c$ phase and the room temperature $Pca2_1$ and $Pmc2_1$ phases commonly discussed in literature. In terms of irreducible representations of the cubic $Pm\bar{3}m$ arisotype, the primary symmetry modes for the $Pc$ phase determined from theory are $M_3^+(a, b, 0)$, $R_4^+(0, 0, c)$ and $\Gamma_4^-(a, b, 0)$, while those for the $Cm$ phase are $M_3^+(a, b, c)$, and $\Gamma_4^-(a, b, 0)$. The $\Gamma$ point mode that drives the ferroelectricity in both phases predominantly involves oxygen displacements and negligible contributions from the cations, consistent with our experimental measurements (see Supplementary Text 2). Both phases also show a $M_5^-$ mode that drives the anti-polar displacements of the cations as depicted in Fig. 3(e) and Extended Data Fig. 6.

Figure 4(f) shows the energy of the newly identified phases from theory along with three previously identified low-energy phases of NaNbO$_3$[12] as a function of in-plane lattice constant and epitaxial strain percentage. The energy of the *Pmc2$_1$* phase at +1.78% strain is the minima for the range of lattice constants used in the calculation and is used as the reference for zero energy. For the in-plane lattice parameter corresponding to epitaxial growth on DyScO$_3$ (marked with a red arrow), all five phases shown in the plot are remarkably close in energy with the maximum energy difference as low as 13.9 meV per formula unit, indicating the likely co-existence of these competing ground states in the system and the need for a local probe for characterization. Further details about the octahedral tilt magnitude and polarization of the *Pc* and *Cm* phases are given in Extended Data Fig. 10.

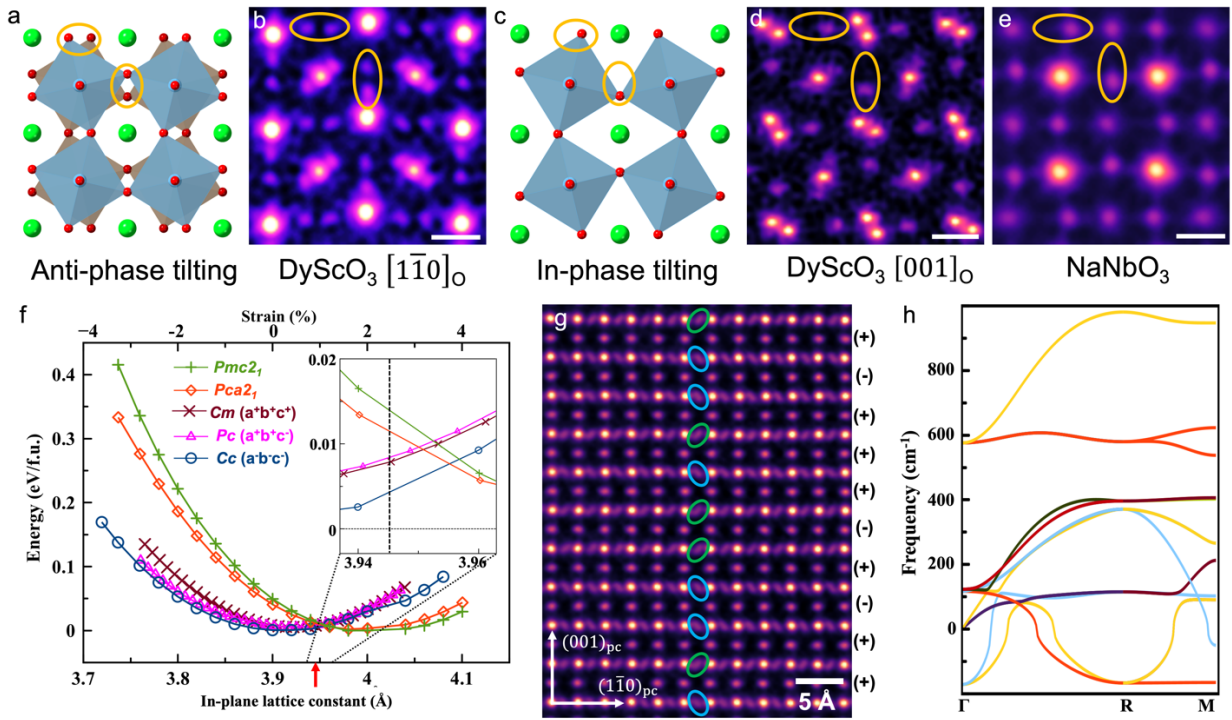

**Fig. 4 | Novel octahedral tilt patterns confirmed with DFT and direct visual evidence of a random tilt pattern.** (*a, c*) *Schematic showing (a) anti-phase and (c) in-phase OOR patterns when viewed in projection along the corresponding axis. The anti-phase/in-phase OOR projections can be differentiated by the presence of double/single oxygen columns as labelled with orange ovals. (**b, d**), MEP images of 2x2 pseudo-cubic unit cells of $DyScO_3$ substrate along the (**b**)$[1\bar{1}0]_O$ zone axis showing an anti-phase OOR pattern with double oxygen columns, and along the (**d**) $[001]_O$ zone axis showing an in-phase OOR pattern with a single oxygen column. (**e**) MEP image of $NaNbO_3$ along an in-plane pseudo-cubic axis, also showing a single oxygen column indicating that the OOR pattern is in-phase. Scale bars in (b, d, e) are 2 Å. (**f**) DFT results showing the energy of newly identified monoclinic phases Pc ($a^+b^+c^-$) and Cm ($a^+b^+c^+$) (with respect to the energy of the $Pmc2_1$ phase at +1.78% strain which is the energy minima in the calculation) as a function of in-plane lattice constant along with some previously identified phases of $NaNbO_3$. For the in-plane lattice constant of $DyScO_3$ substrate relevant in this study and marked with a red arrow, all the different phases of $NaNbO_3$ are very close in energy as highlighted in the inset image. **g**, Ptychographic reconstruction of a different $NaNbO_3$ phase along the $[110]_{pc}$ direction showing a zig/zag pattern for the oxygen atoms marked by blue/green ovals. Along the growth direction, the zig/zag pattern follows a random sequence. A switch between zig-zag or zag-zig is denoted using (+) sign and a recurrence of zig or zag is denoted using (-) sign and is mapped out in the right side of the image. The +/- sign corresponds to a local in-phase/anti-phase tilting between neighbouring oxygen octahedra along the growth direction. **h**, Phonon band structure for the cubic phase of $NaNbO_3$ calculated using DFT shows a flat phonon branch between the M (in-phase) and R (out-of-phase) points of the Brillouin zone.*

## Random octahedral tilt patterns

In addition to the new phase with in-phase tilting mentioned above, from MEP reconstructions, we also identified a structure that exhibits anti-phase tilting along both in-plane directions when imaged along the $[110]_{pc}$ zone axis and is shown in Fig. 4(g). The spatial distribution of phases with in-phase and anti-phase OOR patterns are shown in Supplementary Fig. 5. In terms of the octahedral tilt pattern along the in-plane directions, this structure is like the *Cc* phase ($a^-a^-c^-$), which is the lowest energy structure in DFT calculations for the strain condition corresponding to epitaxial growth on $DyScO_3$ as shown in the inset in Fig. 4(f). The oxygen occupancy along this zone axis takes the form of either a 'zig' or a 'zag' shape (marked with a blue and green oval respectively) for each (001) plane perpendicular to the c-axis. Such a zig/zag configuration is consistent only with the oxygen octahedra having an anti-phase tilting along both in-plane directions; further details can be found in Supplementary Fig. 6. Moreover, the pattern of the zig/zag can be directly mapped to the tilt pattern along the c-axis, with a transition from zig-zag or zag-zig indicating an in-phase tilt (+) and a recurrence of the zig or zag indicating an anti-phase tilt (-) pattern between neighboring pseudo-cubic unit cells along the c-direction. By mapping the tilt pattern from the reconstruction shown in Fig. 4(g), the random nature of the tilt pattern (…+-+++-+-++-…) along the c-axis is evident (see Supplementary Fig. 7. for a larger field of view to exclude the possibility of a large period ordering). Although previous studies have identified incommensurate phases for $NaNbO_3$ using electron diffraction[48] and Raman scattering[7], these MEP reconstructions provide direct visual evidence of the random nature of the tilt patterns in real space for the first time and their spatial extent and transition lengths. These random phases exist as 'nanotwins'[49,50] and originate from the flat nature of the phonon branch between the M (in-phase tilting) and R (anti-phase tilting) points of the Brillouin zone as shown in Fig. 4(h) for the cubic phase of $NaNbO_3$ (bands are present at imaginary frequencies because the cubic phase is not stable at 0 K). Similar flat bands can also be seen in phonon band structures calculated for the dynamically stable cubic phase at higher temperature and *Cc* phase as shown in Supplementary Fig. 8. The flat phonon branch indicates equivalent energies for having a pure in-phase tilt or anti-phase tilt configuration or any intermediate configuration, which explains the random pattern observed experimentally.

## Length-scale for templating octahedral tilts

Characterization of the atomic structure across the interface gives vital clues about the length scale over which the transition from interfacial to bulk behavior occurs in the $NaNbO_3$/$DyScO_3$ heterostructure, and MEP offers the opportunity to study interface properties at a precision and resolution that was not previously achievable. Determining such length scales is especially valuable in the context of the growing interest in octahedral tilt engineering, where octahedral tilts of the substrate are templated onto the film to generate emergent properties[51–54]. Figure 5(a) and (b) show the MEP reconstructions of the NaNbO$_3$/DyScO$_3$ interface along the $[001]_O$ and $[1\bar{1}0]_O$ zone axes of the substrate, respectively. Along the $[001]_O$ zone axis, both the NaNbO$_3$ film and the DyScO$_3$ substrate show in-phase tilting and a coherent interface exists with respect to OORs as shown in Fig. 5(a). Along the $[1\bar{1}0]_O$ axis, DyScO$_3$ exhibits an anti-phase tilting of the oxygen octahedra in contrast to the in-phase tilting of NaNbO$_3$ away from the interface as shown in Fig. 5(b). In Fig. 5(c), we map out this transition from the pure anti-phase tilting in DyScO$_3$ to the pure in-phase tilt in NaNbO$_3$ using the ellipticity of the oxygen columns and the octahedral tilt angle as a metric for quantification. As one moves away from the interface, the ellipticity of oxygen columns and the octahedral tilt angle slowly reach an asymptote over a length scale of around ten unit cells (~ 4 nm). This analysis shows that the upper limit of film thickness at which octahedral tilt engineering could be effectively applied to sodium niobate thin films is around 4 nm. Beyond this thickness, the oxygen octahedra will revert to their "bulk-like" configuration.

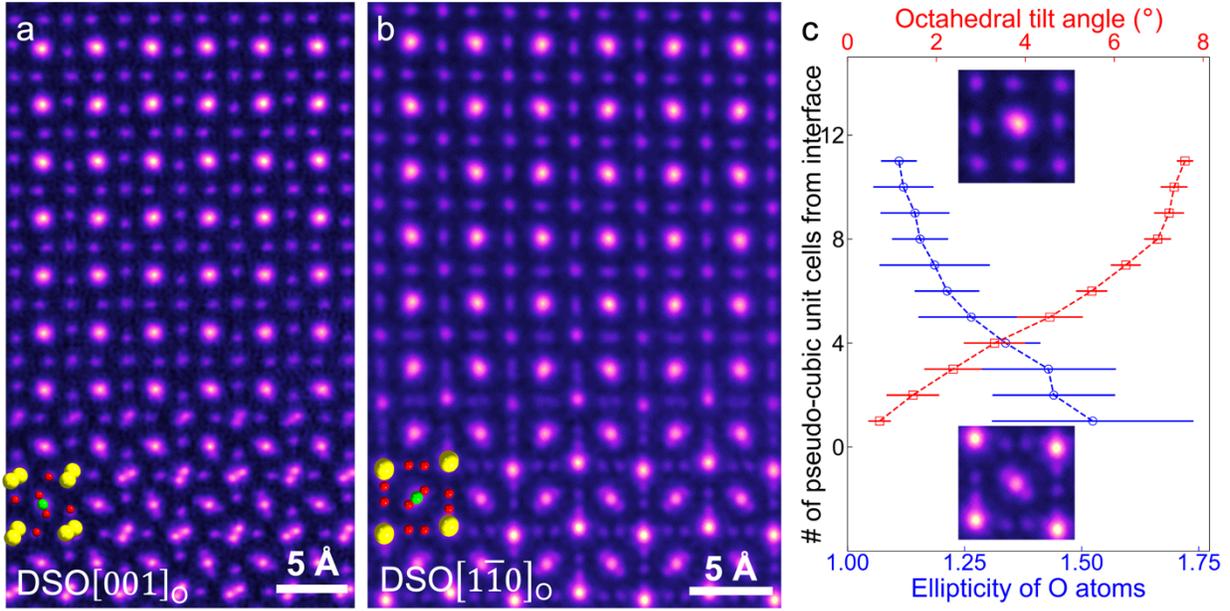

**Fig. 5 | Ptychographic imaging and mapping of octahedral tilts across the interface.** *a-b, Ptychographic reconstructions of the NaNbO$_3$/DyScO$_3$ interface along the $[001]_o$ and $[1\bar{1}0]_o$ directions of the DyScO$_3$ substrate. For **a**, both film and substrate have an in-phase octahedral tilt pattern whereas for **b**, the substrate has an out-of-phase tilt pattern, and the film has an in-phase tilt pattern. **c**, Mapping of the octahedral tilt angle and the ellipticity of the oxygen columns across the interface shown in (**b**) with standard deviation as the error bar. Both quantities reach an asymptote over a length scale of around ten unit cells.*

In summary, we have demonstrated the critical role of oxygen displacements in understanding the microscopic distortions that lead to the macroscopic polarization in a new monoclinic phase of NaNbO$_3$ that is stabilized under epitaxial tensile strain. This finding suggests the potential existence of a broader class of ferroelectric perovskites where polarization is microscopically dominated by anion displacements relative to the cation sublattices, which are not directly coupled to octahedral rotations, and where cation-cation displacements are either minimal or antiferroelectric. This insight also prompts a re-evaluation of search criteria that might otherwise exclude candidates for ferroelectricity. We show how conventional STEM techniques fail to capture the critical structural details in the sample due to electron channeling artifacts or depth-dependent structural variations and how we addressed them using MEP. We also provide direct visual evidence of the random tilt pattern of the oxygen octahedra arising from the flat nature of the phonon branch between points in the Brillouin zone that drive in-phase and anti-phase tilts of the octahedra. Further, by analyzing the propagation of oxygen octahedral tilts across the interface, we identify upper limits on the film thickness for templating octahedral tilt patterns from the substrate.

**Methods**

Thin Film Growth

The 58 nm-thick $NaNbO_3$ films were synthesized on (110)-oriented single-crystal $DyScO_3$ substrates by pulsed laser deposition. The films were grown at a substrate temperature of 600 ºC and an oxygen partial pressure of 200 mtorr. The laser had a fluence of 1.7 J/cm$^2$, a repetition rate of 2 Hz, and an imaged spot size of 4.96 mm$^2$. After the growth, the chamber pressure was raised to 1.5 torr $O_2$ during cooling, with a cool down rate of 5 ºC/min.

Piezoresponse Force Microscopy

The PFM measurements were performed with a Cypher AFM from Asylum research in Vector PFM mode to allow for the simultaneous mapping of both the in-plane and out-of-plane domain structures of the $NaNbO_3$ thin films. The tips used for measurement were Ir/Pt-coated conductive tips with a force constant of approximately 2.8 N m$^{-1}$ (Nanosensor, PPP-EFM). The samples were rotated by 45º with respect to the tip to allow for maximum in-plane signal from the $(110)_{pc}$ oriented domains.

X-Ray Diffraction

The 2D RSM measurements were performed with a PANalytical Empyrean with a monochromated Cu $K_{α1}$ ($λ$ = 1.5406 Å) source. Synchrotron 3D RSM measurements (and the subsequent reciprocal plane slices and line scan cuts) were performed with a five-circle Huber diffractometer with Chi geometry at the 12-ID-D beamline sector of the Advanced Photon Source (APS) at the Argonne National Laboratory. The X-ray beam had energy of 20 keV, wavelength of $λ$ = 0.6199 Å, total flux of 2 × $10^{12}$ photons per second, and a profile of 50 μm ($V$) × 50 μm ($H$). The X-ray energy was manipulated and set using a Si (111) double crystal monochromator with resolution $ΔE/E = 1 × 10^{-4}$ and the beam was vertically de-magnified using 1D focusing compound refractive lenses. A Pilatus 100K 2D area detector was used to record 2D images of the scattering signals which then had geometric corrections applied to them, with data processing being performed using RSMap3D software developed by APS.

Scanning Transmission Electron Microscopy (STEM)

The STEM samples were prepared using a Thermo Fisher Helios G4 UX focused ion beam using the standard lift-out method. The ADF and iDPC images were acquired on an aberration-corrected Thermo Fisher Spectra 300 X-CFEG STEM operated at 300 kV, with a semi-convergence angle of 30 mrad, while the ABF images were acquired on an aberration-corrected FEI Titan Themis at an accelerating voltage of 300 kV with a semi-convergence angle of 21.4 mrad.

Multislice ptychography

The 4D-STEM datasets for multislice ptychography were acquired on an EMPAD[55] detector using a Thermo Fisher Spectra 300 X-CFEG STEM operated at 300 kV, with a semi-convergence angle of 30 mrad, an outer collection angle of 53.3 mrad, step-size of 0.42 Å and a probe overfocus of 10 nm. The probe current was 60 pA with a dwell time of 1 ms per scan position, which evaluates to a total dose of around $2 \times 10^6$ electrons/Å$^2$.

Phase retrieval was performed using the maximum-likelihood multislice ptychography algorithm[56] implemented in the fold-slice package[28,57]. The algorithm uses a gradient based approach for position correction and multiple probe modes to account for the partial coherence of the electron beam[58,59]. Optimal parameters (convergence angle, defocus, sample thickness) for the reconstruction were estimated using a Bayesian optimization model using the data error as the objective function[60]. The final reconstructions were carried out for more than 1000 iterations using 6 probe modes and a slice thickness of 0.5-1 nm. The object phase and amplitude for an example dataset along with histograms showing their range are shown in Supplementary Fig. 9. Since each slice in the reconstruction is assumed to correspond to a phase object, the object amplitude for each slice should remain close to unity. Deviations from unity primarily correspond to scattering beyond the outer collection angle of the detector. The different probe modes in the reconstruction are shown in Supplementary Fig. 10.

Measurement of atomic displacements and polarization

The microscopic dipole moment is characterized by measuring the offset between the centroid of the cations and the centroid of the anions for each pseudo-cubic unit cell. We take the pseudo-cubic unit cell centered on the Nb (B-site) as our reference. We first determine the positions of the atomic columns in the reconstructed ptychographic object using 2d-Gaussian fitting implemented in the Atomap python package[61]. For each niobium atom, the centroid of the surrounding 4 sodium atoms and surrounding 4 oxygen atoms are calculated. The two dipole moments are then given by the vector connecting i) centroid of the oxygen atoms to the centroid of the sodium atoms, and ii) centroid of the oxygen atoms to the position of the niobium atom. The above calculations carry a caveat that the two oxygen atoms aligned with the niobium atoms in projection are excluded in the calculation of the centroid. For traditional characterization of ferroelectrics using ADF images, the vector connecting the centroid of one cation sublattice to the other cation sublattice is typically used as a proxy to describe the polarization. From the measured lengths of the dipoles, the dipole moment can be estimated by assigning either formal charges (related to their valence state - +1 for Na, +5 for Nb and, -2 for O) or Born-effective charges (from DFT calculations) to the different atoms. The vector sum of the two resulting dipole moments gives the total dipole moment per unit cell. Dividing the above value by the unit cell volume yields an estimate of the polarization (dipole moment per unit volume).

Density-functional theory

First-principles calculations employing density functional theory (DFT) were conducted on (001) epitaxial films of NaNbO$_3$. The calculations utilized the projector augmented-wave method[62] within the Vienna Ab initio Simulation Package (VASP) code[63], employing the generalized gradient approximation with the Perdew–Burke–Ernzerhof functional for solids[64]. A plane-wave basis set with a kinetic energy cutoff of 550 eV was utilized. The valence states used for the calculations were $2p^63s^1$ for Na, $4s^24p^64d^45s^1$ for Nb, and $2s^22p^4$ for O, respectively.

The misfit strain which we varied from -4% to +4% is defined as:

$$\text{Misfit strain} = \frac{a-a_0}{a_0} \times 100\% \qquad (1)$$

where $a$ represents the in-plane lattice constant of the substrate and $a_0$ is the in-plane lattice constant resulting from the energy minimization of the $C_c$ phase. For modeling perfect epitaxy on a cubic substrate, the strain tensor, in Voigt notation, has three fixed elements during each simulation. They are: $\eta_1 = \eta_2 = \eta_{misfit}$, $\eta_6 = 0$, while $\eta_3$, $\eta_4$, and $\eta_5$ are allowed to relax, along with all internal atomic coordinates, until the Hellmann-Feynman force on on each atom is converged to be less than 0.001eV/Å for each considered misfit strain. The decomposition of the resulting structures from DFT in terms of irreducible representations of the cubic aristotype structure was performed using the ISODISTORT tool.[65,66]

Phonon dispersion calculation

The phonon dispersion for a temperature of 0 K illustrated in Fig. 4(h) is determined through the density functional perturbation theory[67–69] (DFPT) implemented in VASP[63]. These calculations use the generalized gradient approximation with the Perdew–Burke–Ernzerhof functional for solids[64]. The cutoff energy of 600 eV was utilized to obtained high accuracy dispersions. The Monkhorst-Pack grid in the Brillouin zone was set at 4x4x4 (40 atoms). Subsequently, using the obtained data, the force constants were computed within the phonopy[70] program, resulting in the phonon spectrum.

Simulations of STEM images and 4D-STEM data

Simulations shown in Extended Data Fig. 4 were performed using the multislice method in the μSTEM simulation package[71]. The simulations are infinite dose with an acceleration voltage of 300 kV and an aberration-free probe with a semi-convergence angle of 30 mrad. The BF, ABF and ADF images were obtained by integrating the signal in angular ranges of 0-2 mrad, 15-30 mrad and 60-100 mrad respectively. Both the iDPC and multislice-ptychographic images were calculated using the 4D-STEM dataset simulated with a step size of 0.1975 Å.

Nano-beam electron diffraction (NBED)

The NBED patterns used for identifying the spatial distribution of the in-phase and anti-phase OOR regions and calculation of strain maps were acquired in microprobe mode on an EMPAD[55] detector with a semi-convergence angle of ~1 mrad, dwell time of 1 ms per pixel and beam current of 20 pA. Strain maps were calculated using cepstral analysis[72-74]. Segmentation to separate the in-phase and anti-phase OOR regions was done using k-means clustering on the cepstral transformed data.

**Acknowledgements**

The authors dedicate the paper to Lena Kourkoutis, who has always been an inspiration. H.K.P., Y.-T.S. and D.A.M. acknowledge funding from the Department of Defense, Air Force Office of Scientific Research under award FA9550-18-1-0480, which also supported early stages of NaNbO$_3$ film development. Subsequent work at Stanford/SLAC was supported by the Department of Energy, Office of Basic Energy Sciences, Division of Materials Sciences and Engineering, under contract no. DE-AC02-76SF00515. K.P., S.P. and L.B. thank the Grant MURI ETHOS W911NF-21-2-0162 from the Army Research Office (ARO), the Office of Naval Research Grant No. N00014-21-1–2086, and the Vannevar Bush Faculty Fellowship (VBFF) Grant No. N00014-20-1-2834 from the Department of Defense. This research used resources of the Advanced Photon Source, a U.S. Department of Energy (DOE) Office of Science user facility operated for the DOE Office of Science by Argonne National Laboratory under Contract No. DE-AC02-06CH11357. C.Z. was supported by the NSF under Grant # PHY-1549132, the Center for Bright Beams. The electron microscopy studies made use of the Cornell Center for Materials Research (CCMR) facilities supported by NSF MRSEC program (DMR-1719875), NSF MIP (DMR-2039380), NSF-MRI-1429155 and NSF (DMR-1539918). Part of this work made use of the Stanford Nano Shared Facilities (SNSF), supported by the National Science Foundation under award ECCS-2026822. The authors also thank John Grazul, Mariena Silvestry Ramos and Malcolm Thomas for technical support and maintenance of the electron microscopy facilities.

## Author Contributions

H.K., R.X., Y.-T.S., and D.A.M. designed and planned the research. H.K., and Y.-T.S. performed STEM and MEP characterization under the supervision of D.A.M. PLD synthesis of samples and XRD, PFM characterization was done by R.X., K.J.C., and A.K. under the supervision of H.Y.H. H.Z. was involved in the synchrotron experiments for 3D RSM measurements. K.P., and S.P. performed the DFT calculations under the supervision of L.B. C.Z. wrote parameter optimization codes for MEP used in this study. H.K. wrote the manuscript with feedback from all authors.

## Data availability

The experimental datasets are available on Zenodo (DOI:10.5281/zenodo.13787851).

## Code availability

The codes for the ptychographic reconstructions are also available on Zenodo along with the data (DOI:10.5281/zenodo.13787851).

## Competing interests

Cornell University has licensed the EMPAD hardware to Thermo Fisher Scientific.

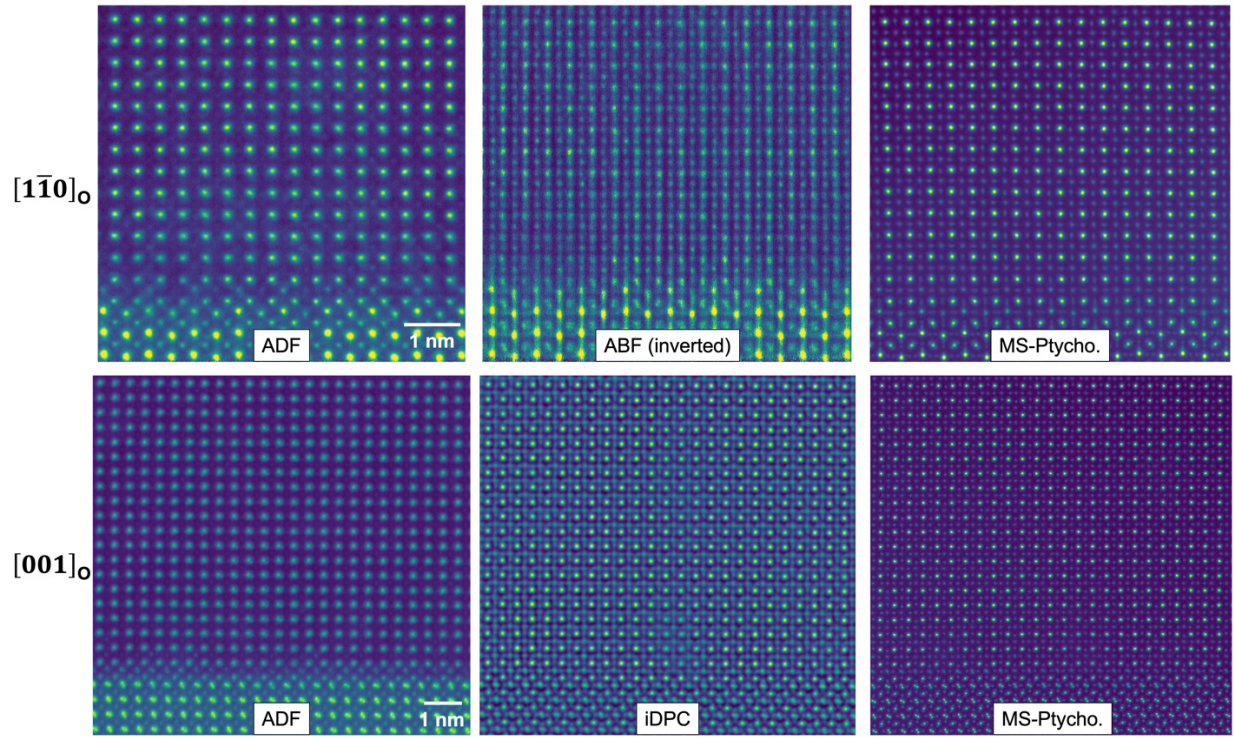

**Extended Data Fig. 1 | Large field of view comparison of multislice electron ptychography with conventional techniques.**
Comparison of images of the NaNbO$_3$/DyScO$_3$ heterostructure acquired with different imaging modalities when viewed along the $[1\bar{1}0]_O$ zone axis (top row) and the $[001]_O$ zone axis (bottom row) of the DyScO$_3$ substrate.

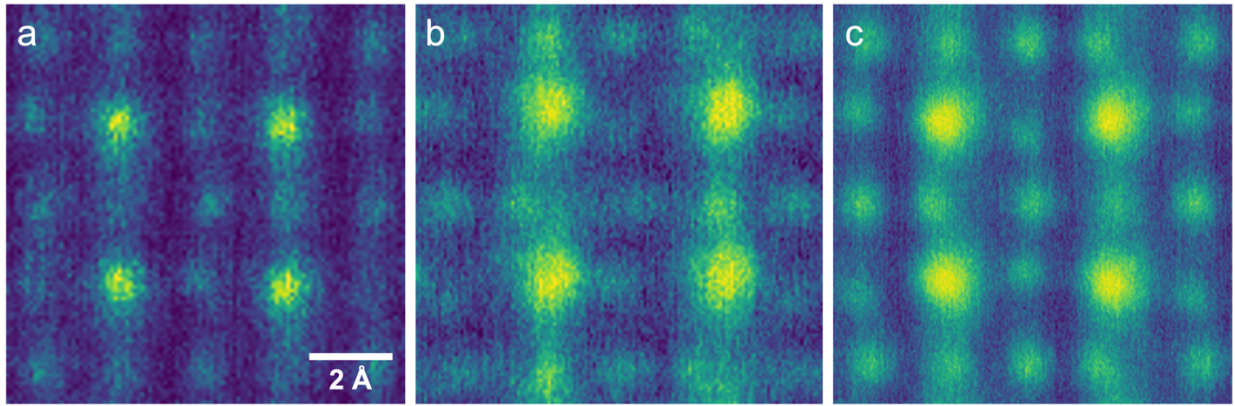

**Extended Data Fig. 2 | Comparing ABF images taken at different sample regions.**
The contrast of ABF images relies heavily on the details of the diffraction conditions and is most sensitive to the sample's surface structure. **(a-c)** show ABF images (inverted contrast) taken at different regions of the $NaNbO_3/DyScO_3$ sample. The images show differences in the positions and contrast of the different atoms, arising predominantly from variations in sample thickness. For very thin regions in the sample, ABF might give qualitatively the correct structure as can be seen in **(c)**. But without prior knowledge of the correct structure, ABF images are unreliable as images taken in different regions of the sample can show very different characteristics.

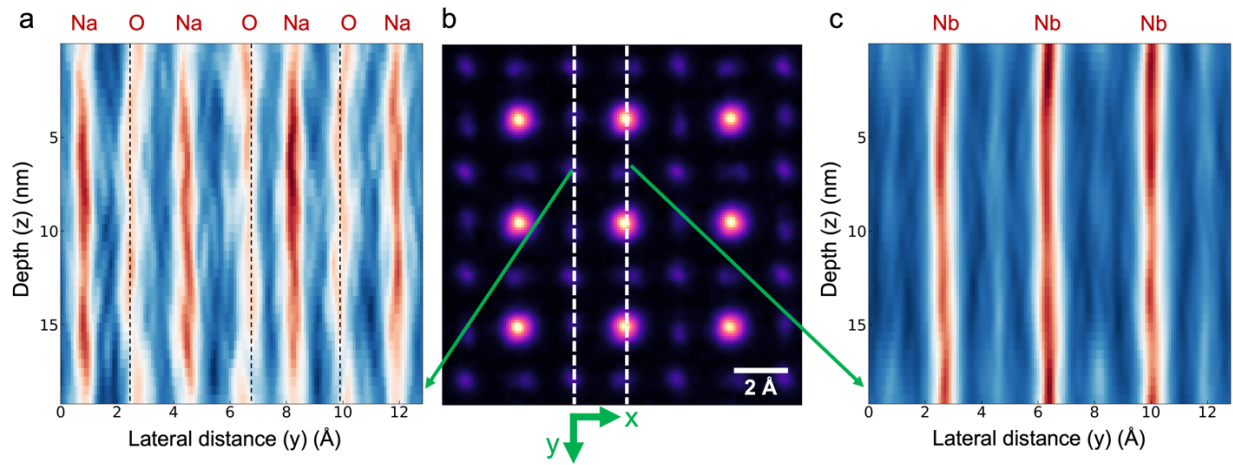

**Extended Data Fig. 3 | Surface relaxations in NaNbO$_3$ grown on SrTiO$_3$ revealed with depth sectioning.**
NaNbO$_3$ thin films grown on SrTiO$_3$ exhibit surface relaxations of a larger magnitude than seen in films grown on DyScO$_3$. **a, c,** show the depth profiles taken along the white dotted lines in **(b)**. The position of the oxygen atom column in the interior of the sample is marked with a black dotted line in **(a)** and shows a clear offset from the position of the atoms at the surface of the sample. The Na atoms exhibit much smaller shifts in atomic position at the surface, whereas the Nb atoms stay at the same position throughout the depth of the sample as shown in **(c)**.

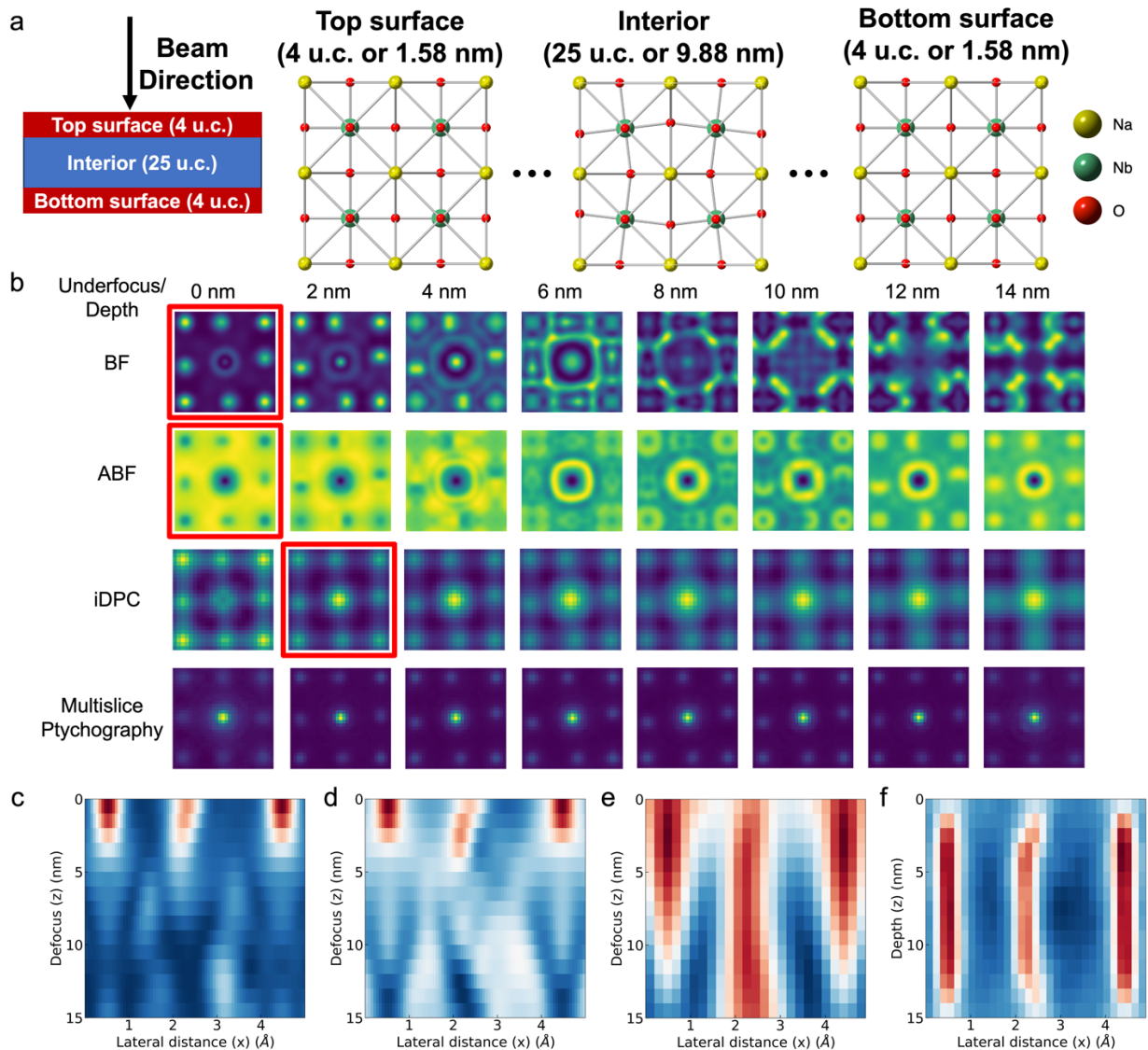

**Extended Data Fig. 4 | Depth sectioning simulations on a modeled sample with surface relaxation**

**a**, Schematic of a simplified model structure for the surface relaxations seen in the NaNbO$_3$ sample – both surface layers have no octahedral rotation, whereas the interior has a 7° octahedral rotation about the pseudo-cubic axis parallel to the beam direction. **b**, shows a simulated through-focal series of BF, ABF, iDPC images and different slices from a multislice ptychographic reconstruction. The simulations are infinite-dose and assume an aberration-free probe. For each imaging modality, the image that appears best visually in the defocus series is highlighted with a red border and corresponds to the beam focused near the top surface. The BF and ABF images undergo contrast reversals and show clear atomic contrast only for a limited range of defocus. The contrast in the best ABF and iDPC image is a closer match to the relaxed surface structure than the interior structure, and hence would produce large discrepancies if used for comparisons with other macroscopic measurements. The multislice ptychographic reconstruction is able to capture the relaxed structure at the two surfaces as well as the interior structure accurately. **c-f**, Depth profiles along the Na-O plane for (c) BF, (d) ABF, (e) iDPC and (f) multislice ptychographic reconstruction. Away from their ideal defocus condition, the BF, ABF and iDPC images do not have the requisite contrast or resolution to unambiguously differentiate the presence of the relaxed surface structure from the bulk structure. Multislice ptychography enables accurate depth sectioning from a single dataset with a clear distinction between the surface and interior structure. We also note that the addition of probe aberrations or small mistilts that are practically unavoidable in an experimental situation will cause severe deterioration in the quality of the BF, ABF and iDPC image. Multislice ptychography can recover the correct probe structure even in the presence of aberrations - an example from an experimental dataset can be seen in Fig. S5. Mistilts will produce tilted atomic columns in the depth sectioning, which can easily be corrected by aligning the different slices in a post-processing step.

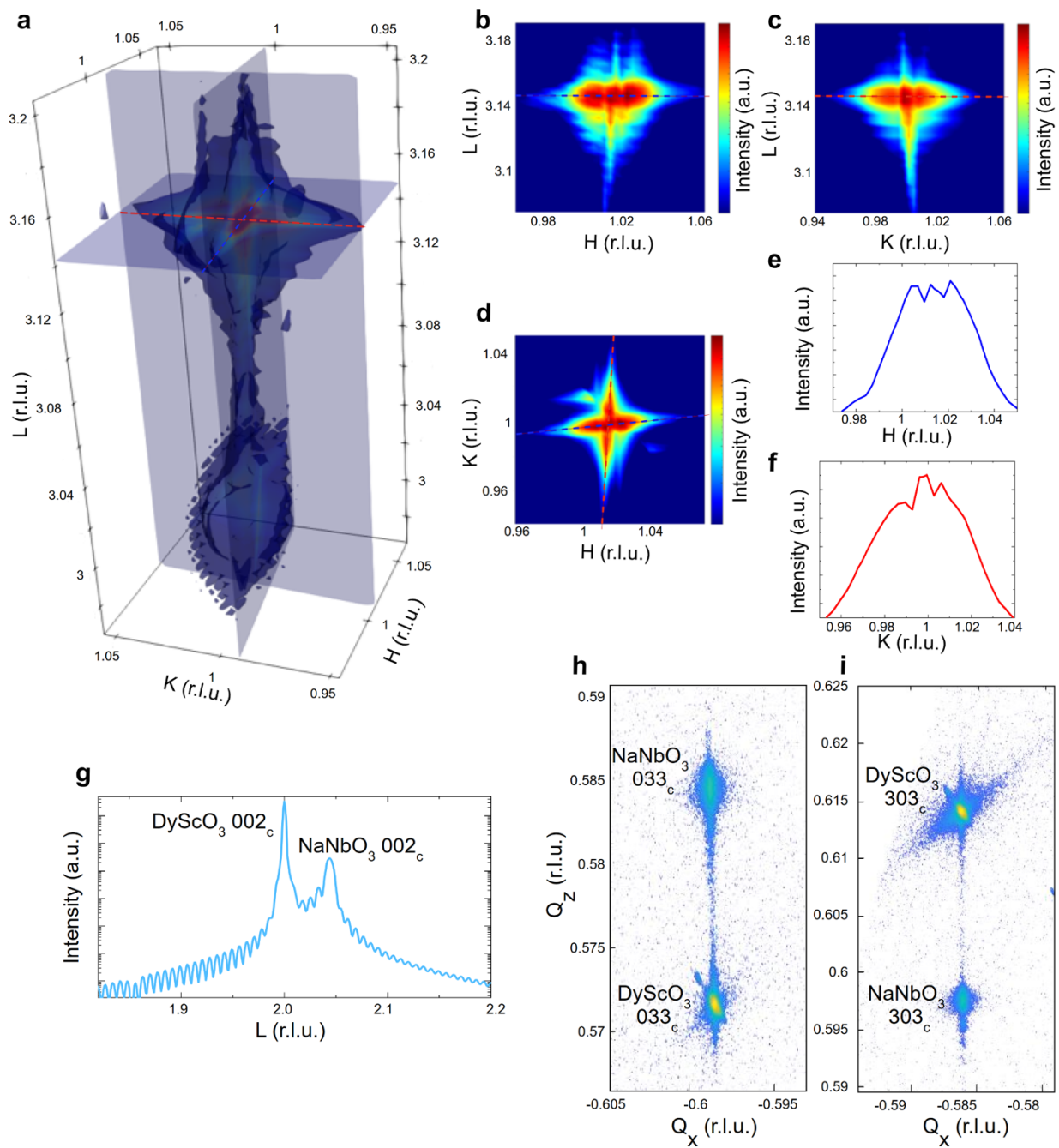

**Extended Data Fig. 5 | Crystalline quality and film strain characterized with reciprocal space mapping and X-ray diffraction.**
Reciprocal space mapping (RSM) and X-ray diffraction (XRD) results. **(a)** Synchrotron-based 3D RSM taken at the (113)$_{pc}$ peak of DyScO$_3$ with corresponding slices along the **(b)** *H-L*, **(c)** *K-L*, and **(d)** *H-K* reciprocal planes. The **H-K** plane was sliced aligned with the NaNbO$_3$ peak as shown in (a) and displays four satellite peaks due to the periodic domain structure, as is commonly seen in ferroelectrics. **(e, f)** Line scan cuts of the *H-K* plane along the dashed lines from (d), which can be used to extract a domain periodicity of 38 nm along *H* and 44 nm along *K*. **(g)** Synchrotron-based 002-line scan measurement showing strong and clear Laue oscillations, confirming the quality of our NaNbO$_3$ films. **(h, i)** 2D RSM taken at the (033)$_{pc}$ and (303)$_{pc}$ peaks, respectively, showing that the film is coherently strained along both in-plane directions. From a refinement of these results, we extract lattice parameters for NaNbO$_3$ of $a = 5.519$ Å, $b = 5.516$ Å, and $c = 15.774$ Å and unit cell angles of $\alpha = 90°$, $\beta = 90°$, and $\gamma = 91.3°$. We can thus calculate the strain for our NaNbO$_3$ film to be 0.8% tensile along $ab_{DSO}$ and 1.6% tensile along $c_{DSO}$.

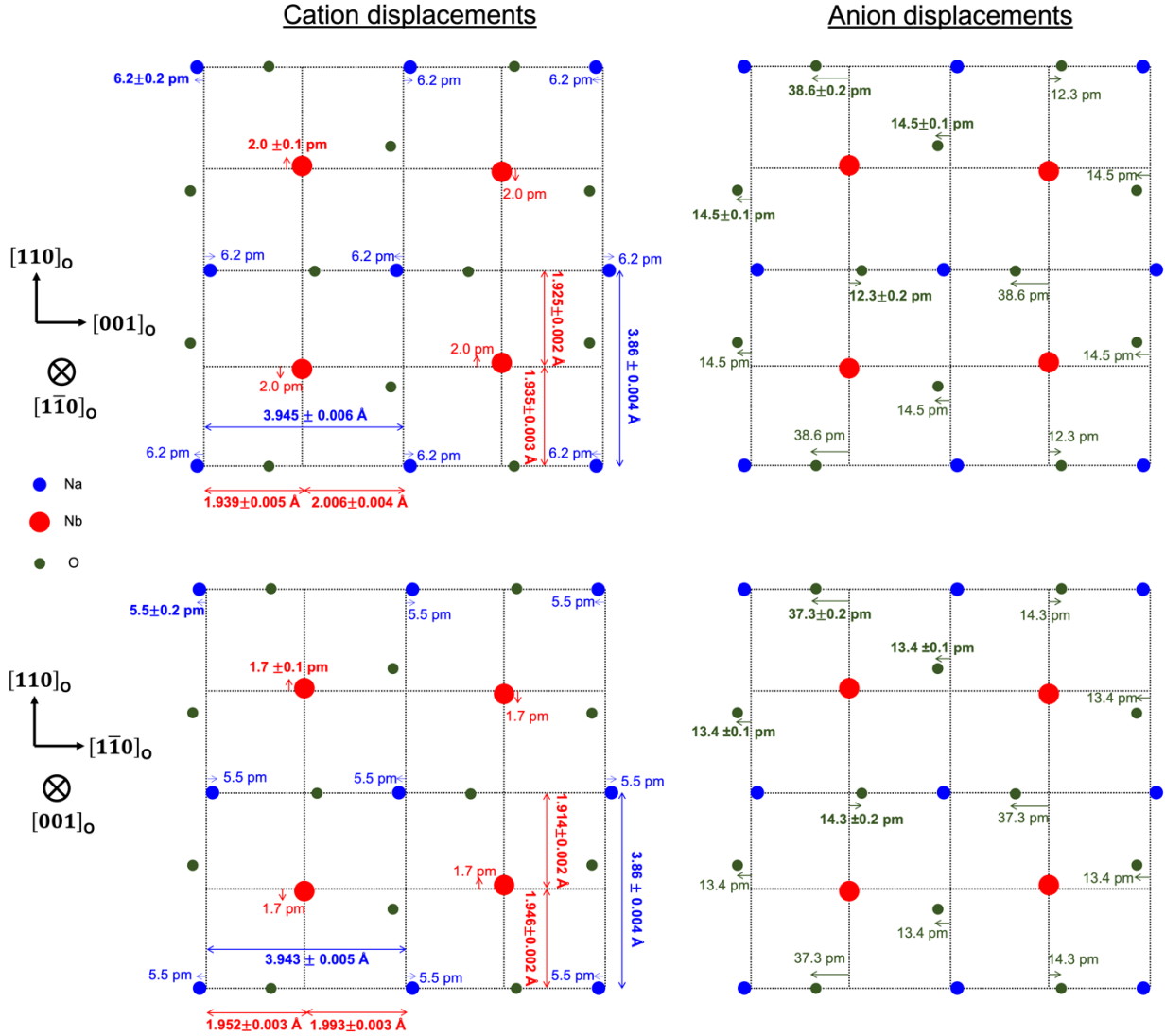

**Extended Data Fig. 6 | Structural distortions in the NaNbO₃ films measured for projections along both in-plane pseudo-cubic axes.**
Schematic showing the major cation and anion displacements in the NaNbO₃ structure for the in-plane $[1\bar{1}0]_O$ (top row) and $[001]_O$ (bottom row) zone axes (directions in reference to the dysprosium scandate substrate). The displacements of the different atoms from the reference planes (black dotted lines) are average of the measured values from the regions shown in Extended Data Fig. 8 (a-c, d-f). The Na atoms show anti-polar displacements along both in-plane pseudo-cubic directions, and the Nb atoms show anti-polar displacements in the growth direction. The net displacement magnitude of the oxygen atoms that displace to the left is much larger than for the net displacement to the right and is the major source of the polarization. The small discrepancies (< 1 pm) in the displacements in the growth direction for the two projection directions might arise from inhomogeneities in different regions of the sample.

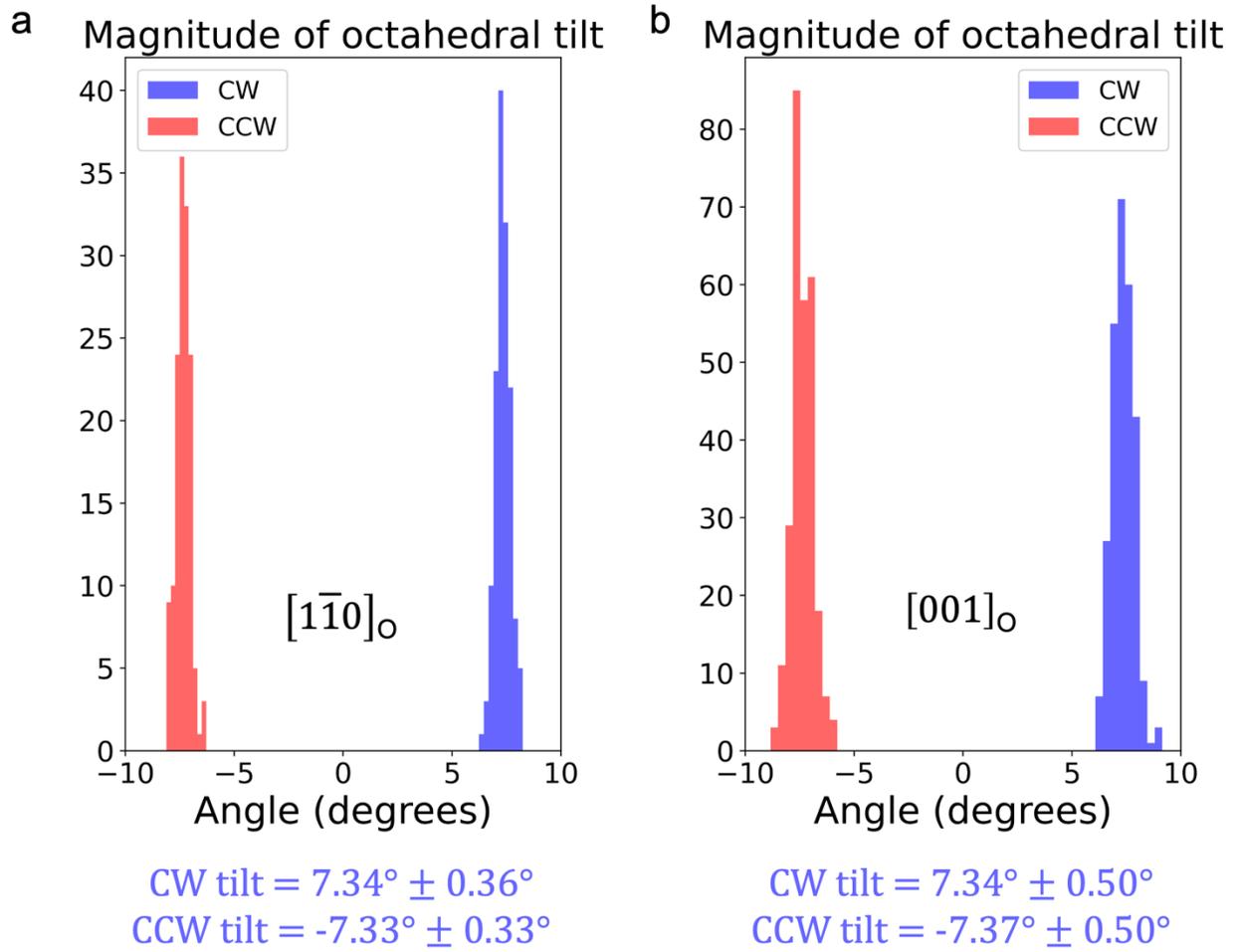

**Extended Data Fig. 7** | Magnitude of oxygen octahedral tilt in the sodium niobate thin film along the **(a)** $[1\bar{1}0]_O$ and **(b)** $[001]_O$ direction of the DSO substrate.

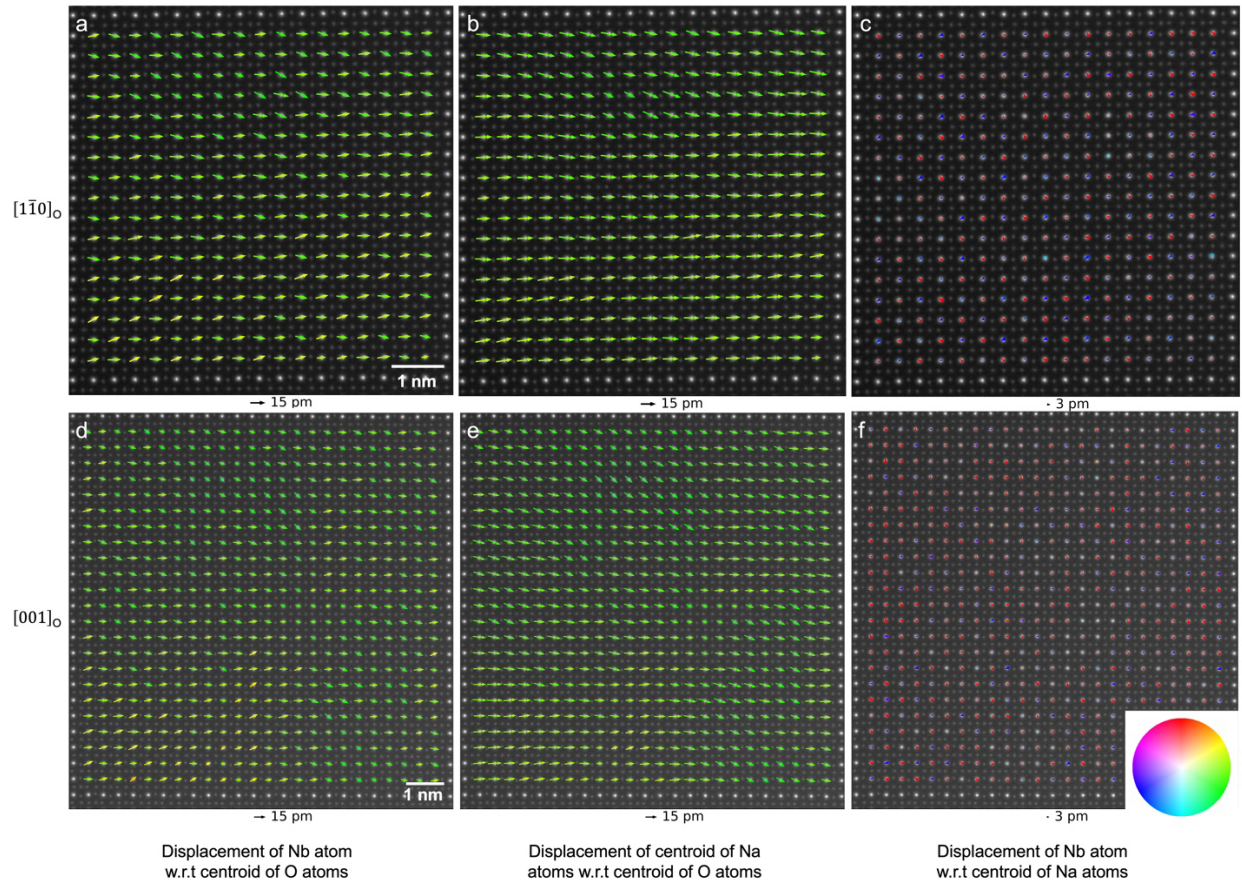

**Extended Data Fig. 8 | Atomic displacement maps along both in-plane pseudo-cubic directions.**
The vector maps show large displacements of the cation centroids with respect to the centroid of the oxygen anions - **(a, d)** for Nb atoms and **(b, e)** for Na atoms, and form the two physical dipoles in the system. The polarization is predominantly in-plane with a small out of plane component. In contrast, the displacement of one cation sublattice with respect to the other **(c, f)** is much smaller and accounts for a very small fraction of the total polarization. The above displacement maps emphasize the importance of accurately measuring the oxygen atom positions for an accurate characterization of the polarization.

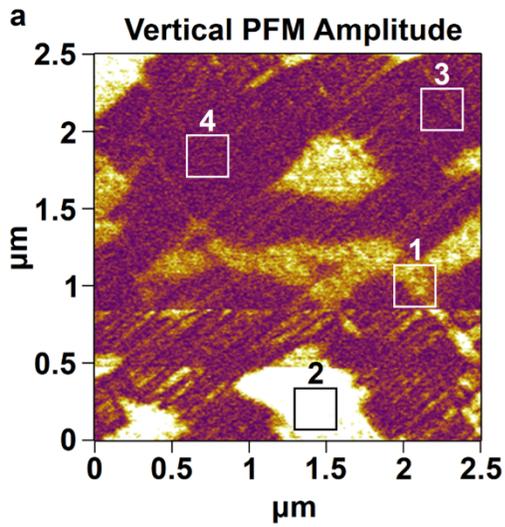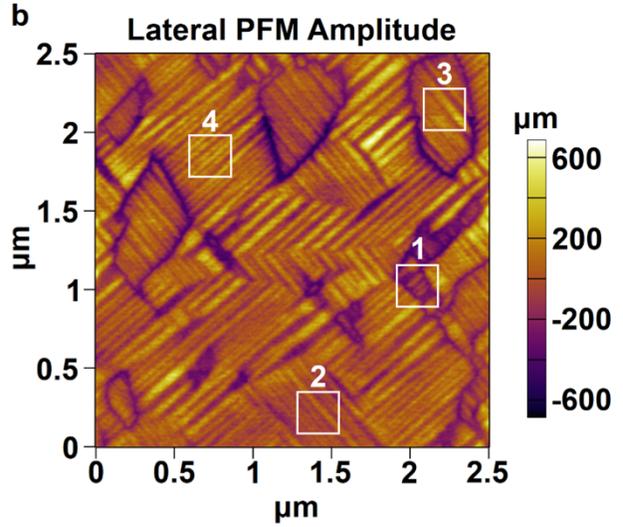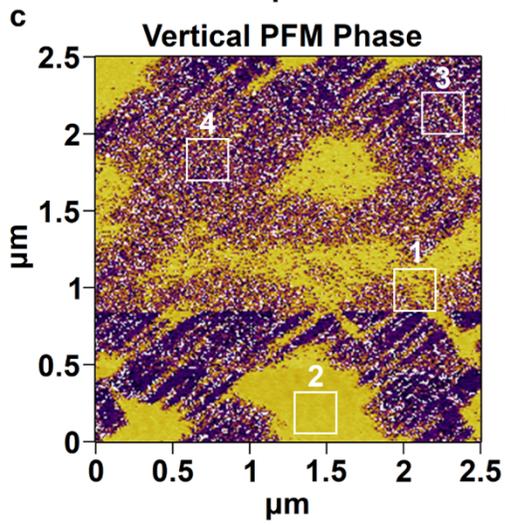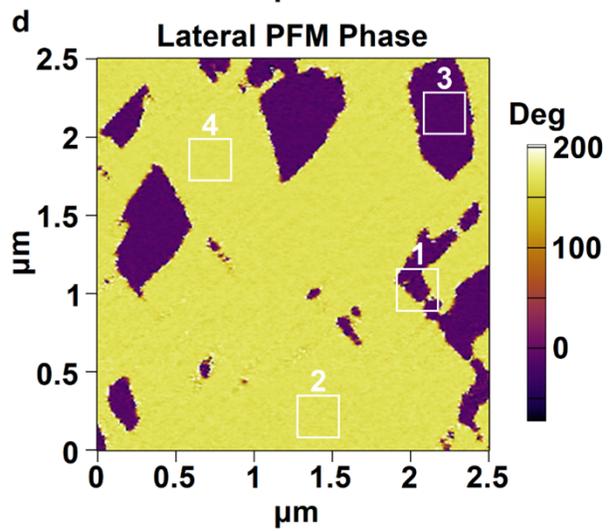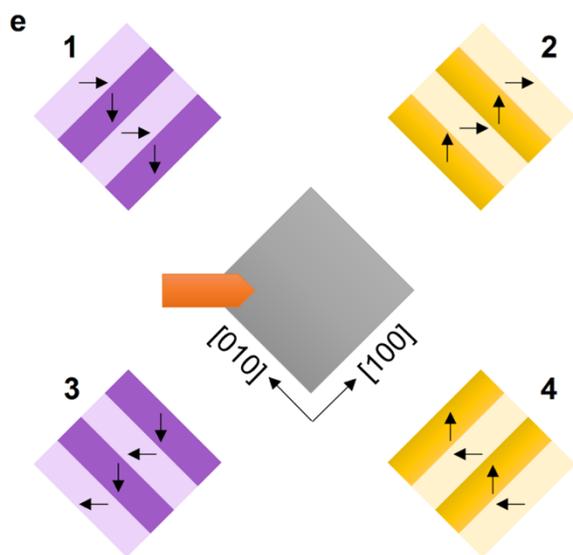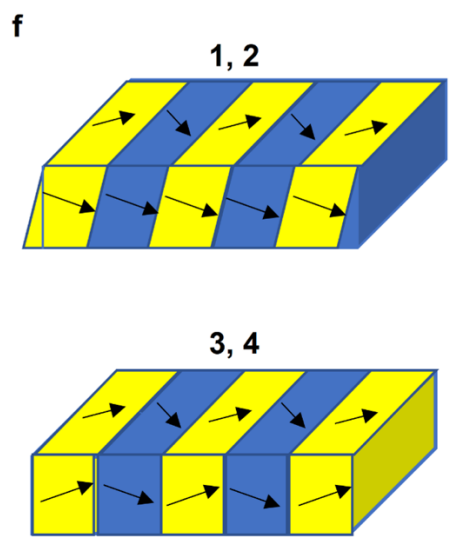

**Extended Data Fig. 9 | Piezoresponse force microscopy results and schematics.**
Piezoresponse force microscopy (PFM) images showing the amplitude **(a, b)** and phase **(c, d)** in the vertical and lateral directions, respectively. The lateral direction displays a much clearer domain pattern with stronger PFM signal when compared to the vertical direction, which matches the ptychography result with polarization predominantly oriented along the $[110]_{pc}$ direction. We can also extract an average domain periodicity of 44 nm along $[100]_{pc}$ and 52 nm along $[010]_{pc}$, which are in rough agreement with the results from X-ray diffraction in Extended Data Fig. 5. Schematics of the domain configurations for the **(e)** lateral and **(f)** vertical measurements are also provided. For these measurements, the sample is rotated by 45º with respect to the tip, as polarizations orthogonal to the tip provide the strongest lateral signal and the polarization is known to be mostly along $[110]_{pc}$. There are four superdomain variants described in these schematics which have been labelled in **(a-d)**. For lateral PFM measurement, the polarization of individual domains will oscillate within a superdomain between pointing along the tip (providing lower amplitude in **b** and shown in pastel colors in **e**) and orthogonal to the tip (providing higher amplitude in **b** and shown in darker colors in **e**). Superdomains 1 and 3 show the same lateral phase because the PFM tip is torqued in the same direction by the orthogonal domains for these regions (the same is also true for superdomains 2 and 4). For the vertical PFM images, we can see that two superdomain variants (3 and 4) have roughly no net polarization since the polarization of neighboring domains will cancel, and thus the vertical PFM amplitude is low in **a**. We note that even in these low amplitude regions, oscillating phases can be observed in **c** as expected from **f**. However, in regions 1 and 2 the vertical polarizations of neighboring domains are aligned, and so the vertical PFM amplitude is higher in **a**, and those regions have a single phase in c.

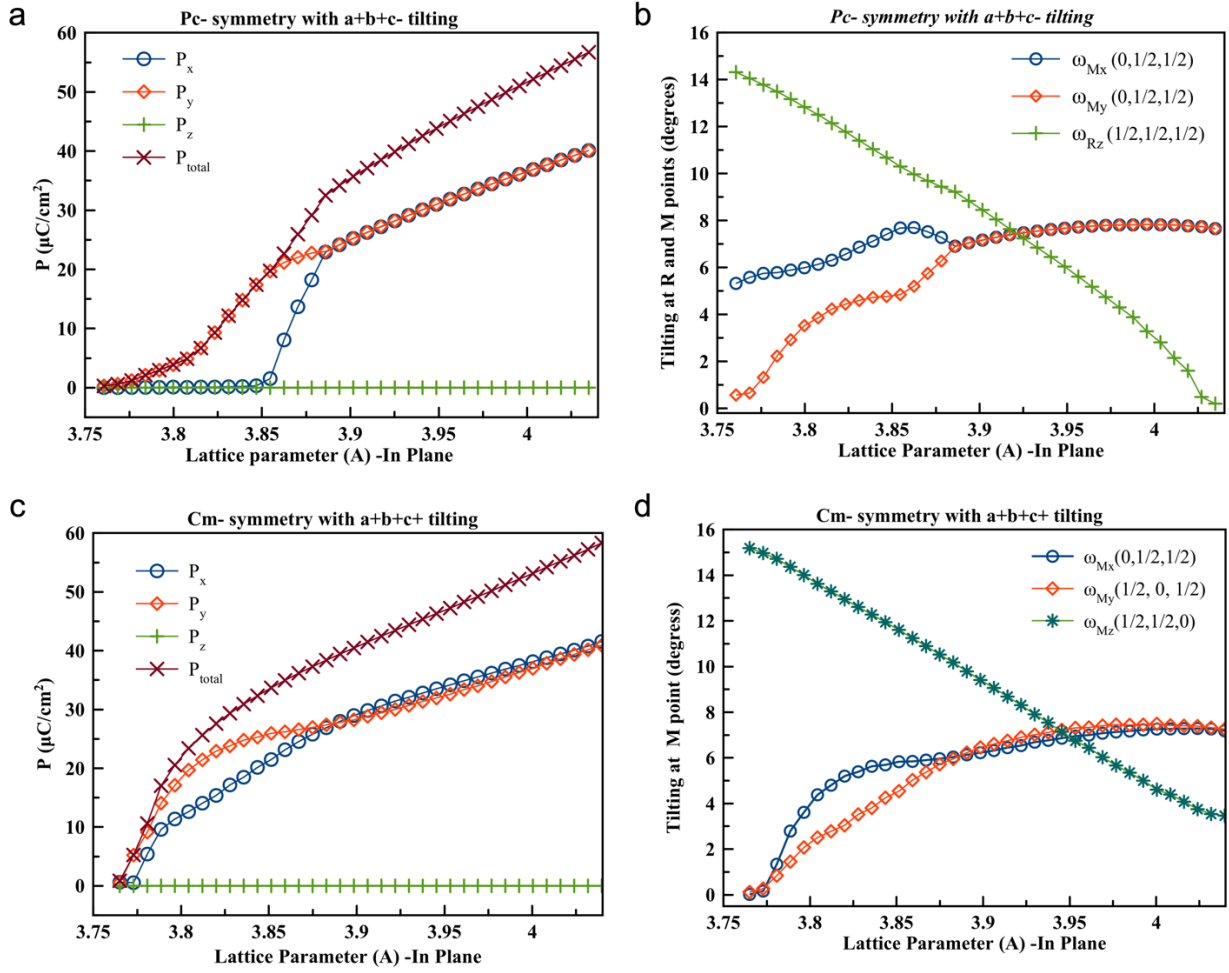

**Extended Data Fig. 10 | Polarization and oxygen octahedral tilt magnitude for *Pc* and *Cm* phases of NaNbO$_3$.**
In-plane lattice parameter dependence of **(a,c)** the polarization and **(b,d)** tilting of oxygen octahedra at R and M point of the first Brillouin zone for the **(a,b)** *Pc* and **(c,d)** *Cm* phases of epitaxial (001) NaNbO$_3$ thin films.

**Supplementary Information for**

**Electron ptychography reveals a ferroelectricity dominated by anion displacements**

Harikrishnan KP,[1] Ruijuan Xu,[2,3,4] Kinnary Patel,[5] Kevin J. Crust,[2,6] Aarushi Khandelwal,[2,3] Chenyu Zhang,[1] Sergey Prosandeev,[5] Hua Zhou,[7] Yu-Tsun Shao,[1,8] Laurent Bellaiche,[5,9] Harold Y. Hwang,[2,3] David A. Muller[1,10,*]

[1.] School of Applied and Engineering Physics, Cornell University, Ithaca, NY, USA.
[2.] Stanford Institute for Materials and Energy Sciences, SLAC National Accelerator Laboratory, Menlo Park, CA, USA.
[3.] Department of Applied Physics, Stanford University, Stanford, CA, USA.
[4.] Department of Materials Science and Engineering, North Carolina State University, Raleigh, NC, USA.
[5.] Smart Ferroic Materials Center, Physics Department and Institute for Nanoscience and Engineering, University of Arkansas, Fayetteville, Arkansas 72701, USA.
[6.] Department of Physics, Stanford University, Stanford, CA, USA.
[7.] X-ray Science Division, Advanced Photon Source, Argonne National Laboratory, Lemont, IL, USA.
[8.] Mork Family Department of Chemical Engineering and Materials Science, University of Southern California, Los Angeles, CA, USA.
[9.] Department of Materials Science and Engineering, Tel Aviv University, Ramat Aviv, Tel Aviv 6997801, Israel.
[10.] Kavli Institute at Cornell for Nanoscale Science, Ithaca, NY, USA.
*Corresponding Author: E-mail: david.a.muller@cornell.edu

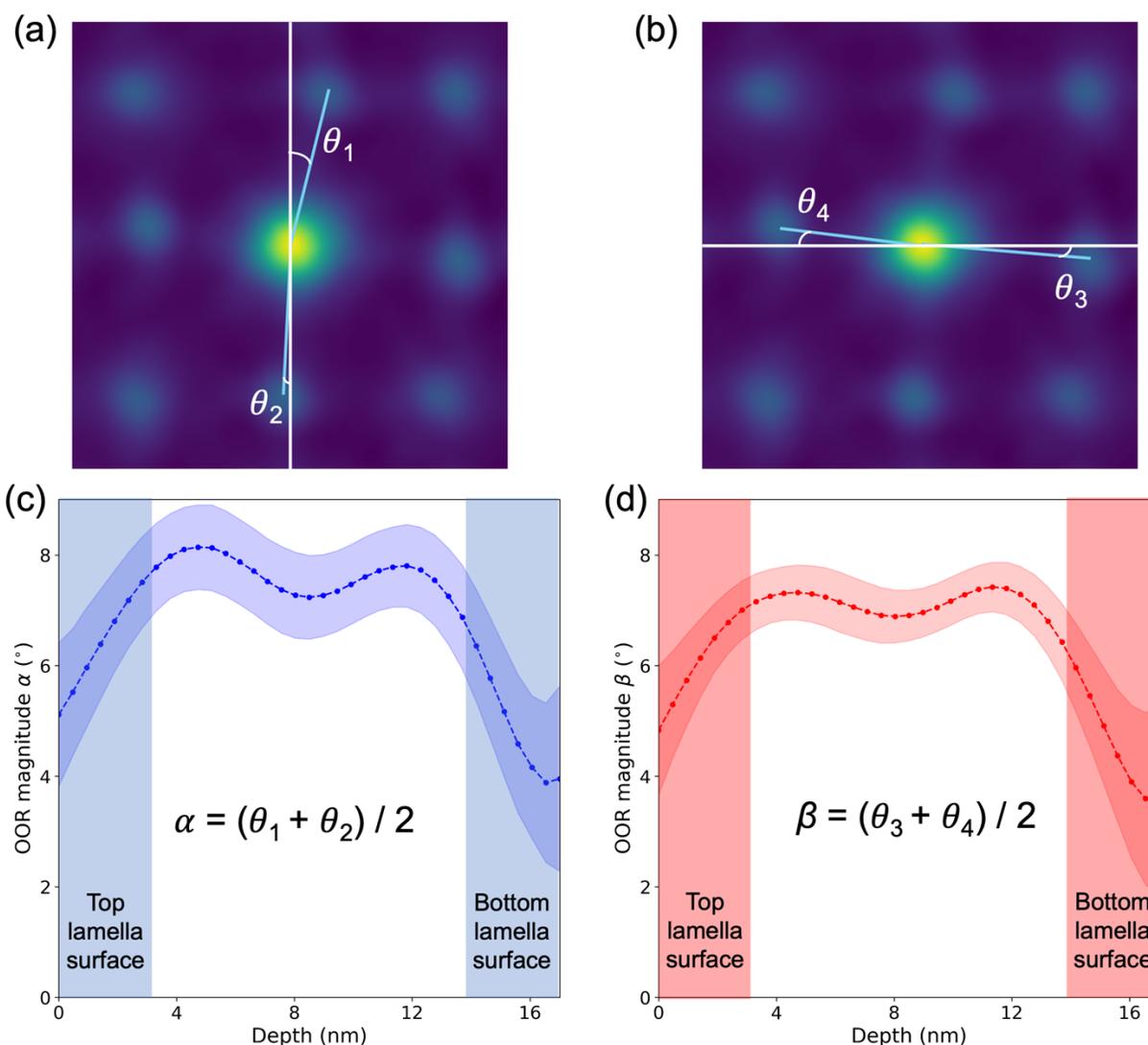

**Fig. S1** | **(a, b)** Schematic showing the different Nb-O bond angles that are used to estimate the magnitude of OOR. **(c, d)** OOR magnitude plotted as a function of depth direction, calculated by quantification of the Nb-O bond angles for individual slices in the MEP reconstruction. The error band is used to indicate the standard deviation in the measurement of the angles. Shaded areas are used to demarcate the top and bottom surfaces of the lamella from the sample interior. The OOR magnitude is suppressed at the two lamella surfaces in comparison to the interior.

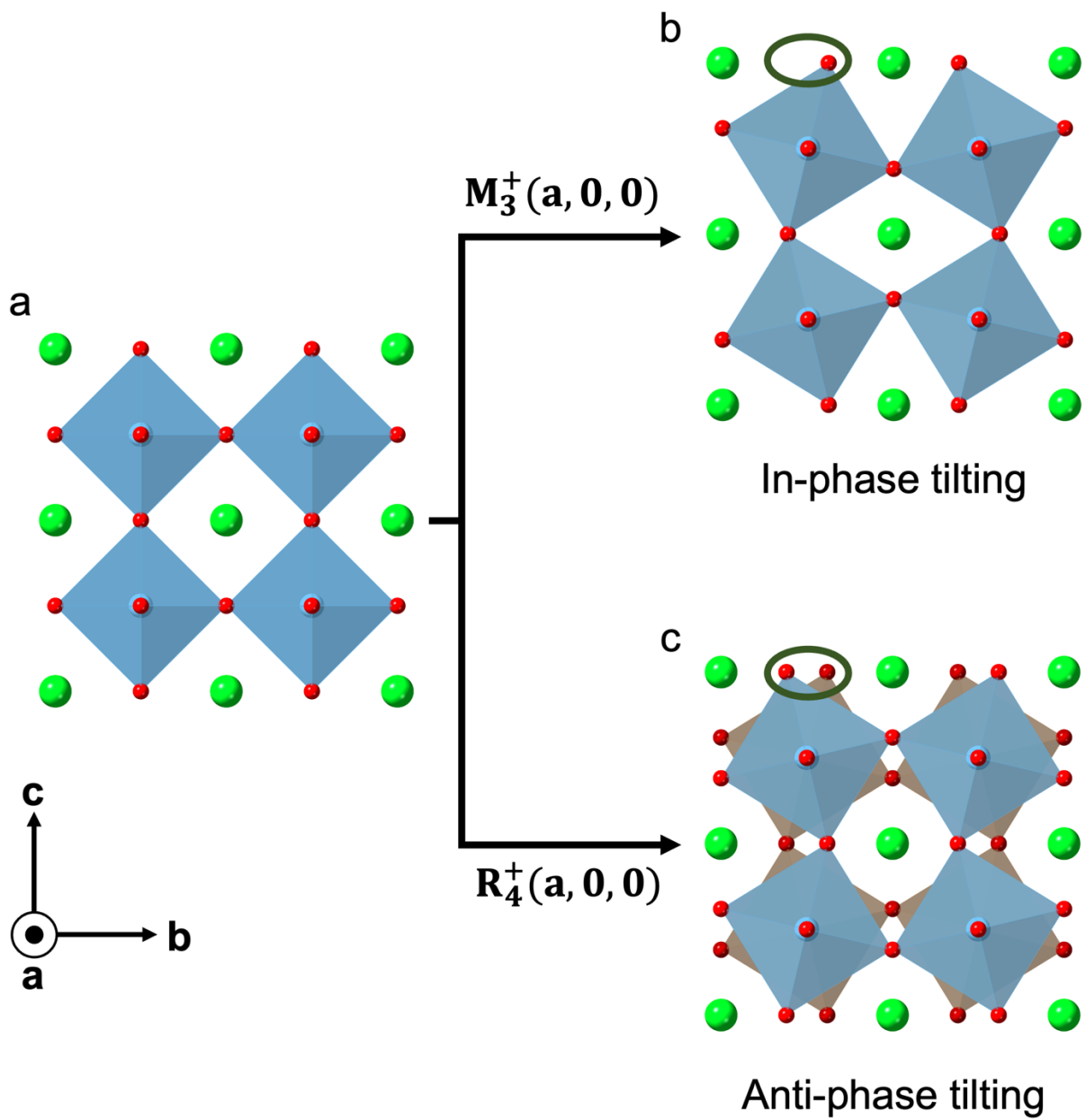

**Fig. S2** | Schematic illustrating the difference between in-phase and anti-phase rotations of the oxygen octahedra. The parent aristotype structure with no octahedral rotations is shown in (a), with green, blue, and red spheres used to indicate the A-site, B-site, and oxygen sites, respectively. The $M_3^+$ mode drives in-phase rotations of the octahedra as shown in (b) where all the octahedra along the rotation axis (a-axis) tilt in the same direction. The $R_4^+$ mode drives anti-phase tilting of the octahedra as shown in (c), where alternate octahedra (shown in different shades) along the rotation axis tilt in opposite directions. In projection, these modes can be differentiated by the presence of a single (in-phase) or split (anti-phase) oxygen sites as highlighted with the dark green ovals.

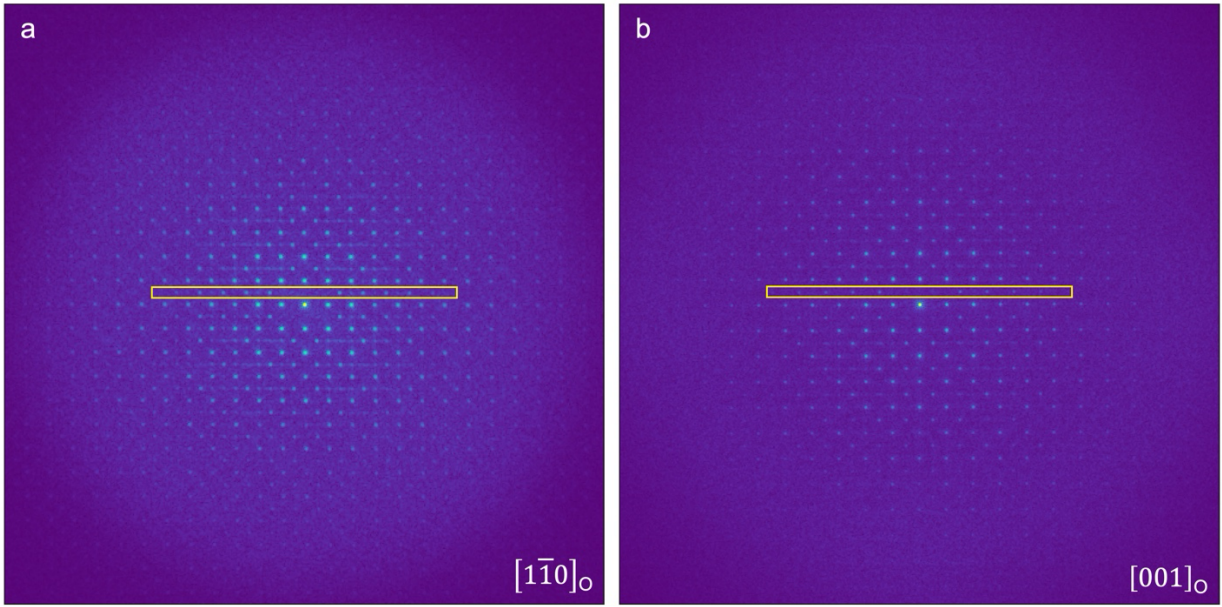

**Fig. S3** | Magnitude of the Fourier transforms of the reconstructed phase images of the sodium niobate thin film projected along the **a**, $[1\bar{1}0]_O$ and **b**, $[001]_O$ directions of the DyScO$_3$ substrate. A row of half-order satellite peaks arising from the oxygen octahedral rotations and anti-polar cation displacements is highlighted with a yellow box in both images. The images are weighted with a power of 0.25 to highlight the superlattice peaks and information transfer at higher frequencies.

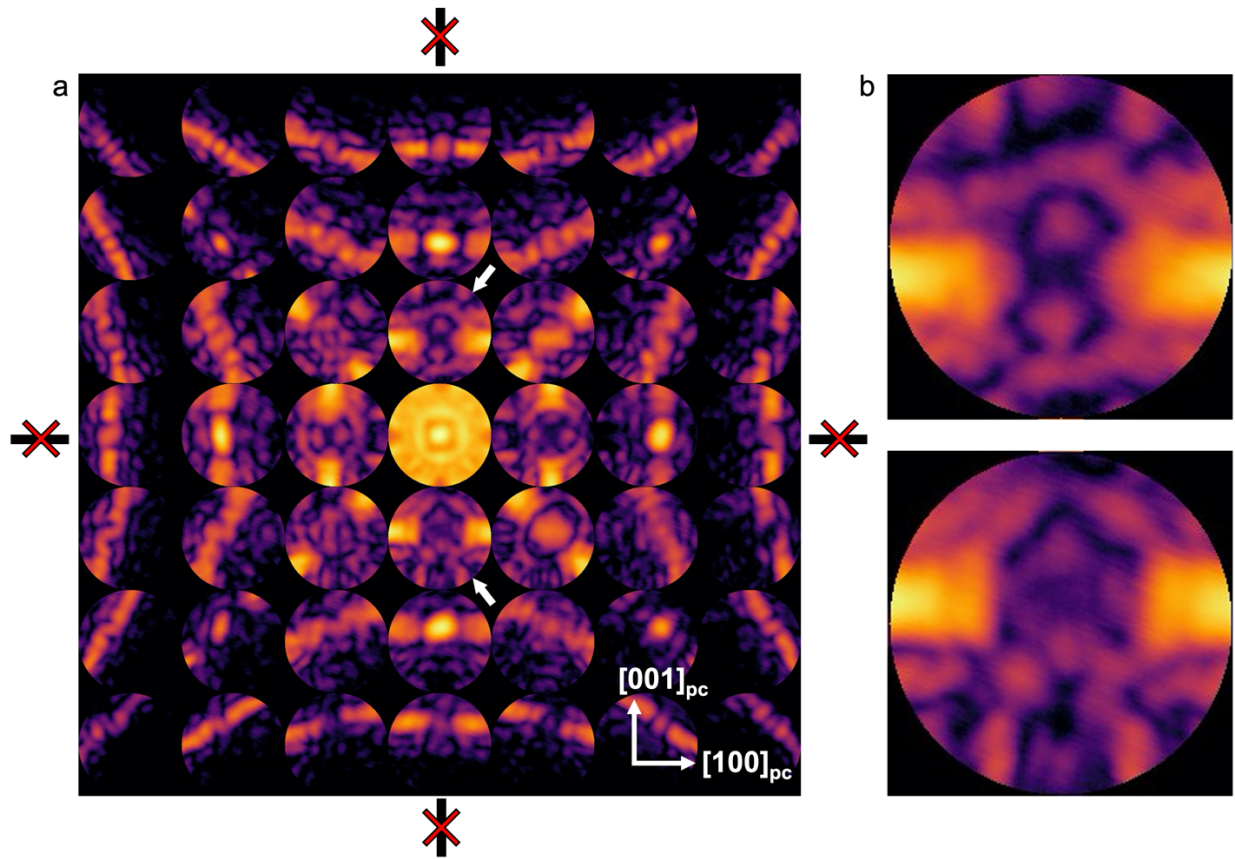

**Fig. S4** | **a**, Large angle rocking beam electron diffraction[1] (LARBED) pattern of the sodium niobate thin film collected with a maximum tilt angle of 1.5 degrees. As expected, the mirror plane parallel to the in-plane $[100]_{pc}$ direction is broken due to the large in-plane polarization. In addition, the mirror plane parallel to the $[001]_{pc}$ direction is also broken due to the presence of a small out-of-plane polarization and is consistent with PFM results. The presence of a net out-of-plane polarization also suggests monoclinic or lower symmetry for the crystal structure. **b**, Magnified images of the $(001)_{pc}$ and $(00\bar{1})_{pc}$ diffraction spots marked with white arrows in (a) show the breaking of the Friedel symmetry due to the out-of-plane polarization.

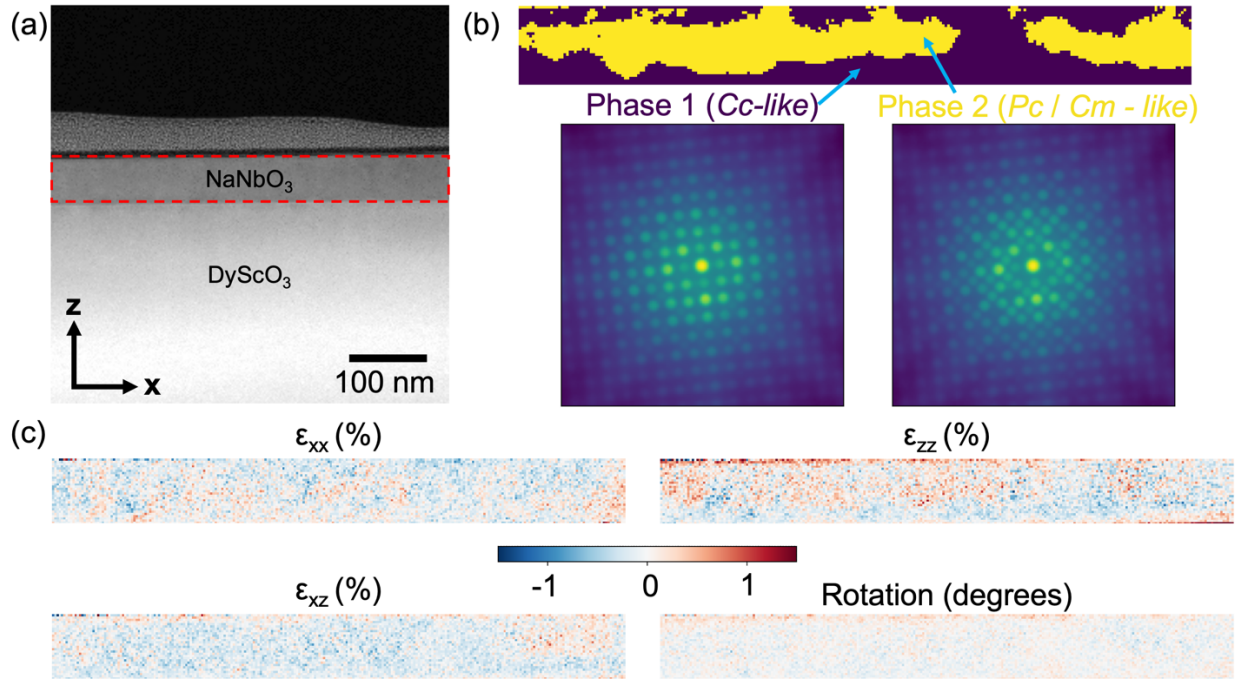

**Fig. S5 | a,** Virtual microprobe ADF image of the heterostructure with the film region labelled with a red dashed outline. **b,** Spatial distribution of phases with anti-phase (Phase 1) and in-phase (Phase 2) OOR marked with blue and yellow color respectively. The phases are distinguished by the absence/presence of half-order satellite peaks in the diffraction pattern. The segmentation of the two phases is done by k-means clustering with cepstral patterns calculated from 4D-STEM data. **c,** Strain maps calculated with the cepstral method from the same 4D-STEM dataset. There is little to no strain contrast across phase boundaries.

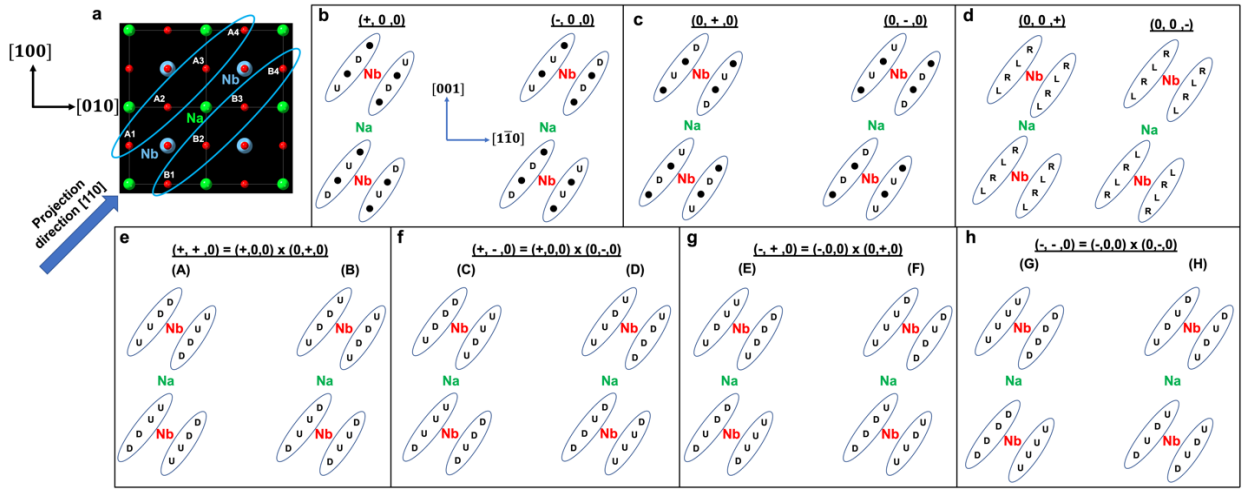

**Fig. S6 | a**, Schematic of an undistorted perovskite – the set of oxygen columns labelled A = {A1,A2,A3,A4} would appear at the same site when projected along the [110] direction. Similarly, the set B = {B1,B2,B3,B4} would all lie on the same site but adjacent to the position of set A and separated by the labelled Nb columns. In **(b-h)**, we look at three consecutive (001) planes down the [110] projection direction which can be used to uniquely determine how the projected structure along $[110]_{pc}$ changes in the presence of octahedral rotations. **(b-d)**, Schematic showing whether the oxygen atoms displace up (U) or down (D) with respect to the $(001)_{pc}$ planes and left (L) or right (R) with respect to the $(1\text{-}10)_{pc}$ planes when viewed down the $[110]_{pc}$ zone axis in response to oxygen octahedral rotations about the (b) $[100]_{pc}$, (c) $[010]_{pc}$, and (d) $[001]_{pc}$ axes. **(e-h)** Schematic of the displacements of the oxygen atoms in response to different combinations of oxygen octahedral rotations about the two in-plane $[100]_{pc}$ and $[010]_{pc}$ axes. There are two combinations in each case depending on whether the upper plane for rotation about one axis is matched with the upper or lower plane for rotation about the other axis. In the ptychography reconstructions along $[110]_{pc}$ axis shown in Fig. 4(g), each $(001)_{pc}$ plane has either UL + DR (oxygen atoms displaced up and to the left and oxygen atoms displaced down and to the right) or UR + DL (oxygen atoms displaced up and to the right and oxygen atoms displaced down and to the left) configurations, giving rise to a zig or a zag. There are no planes where oxygen atoms that have displaced up/down have both left and right displacements and vice-versa. By combining (d) with (e-h), one can see that such a configuration is possible only for (-,-,+) or (-,-,-) tilt configurations. Moreover, on moving between neighboring $(001)_{pc}$ planes, a transition from zig-zag or vice-versa is consistent only with (-,-,+) while a repetition of zig or zag is consistent only with (-,-,-). Mapping the zigs/zags in Fig. 4(g) and Fig. S7 reveals the incommensurate nature of the tilt pattern along the $[001]_{pc}$ axis.

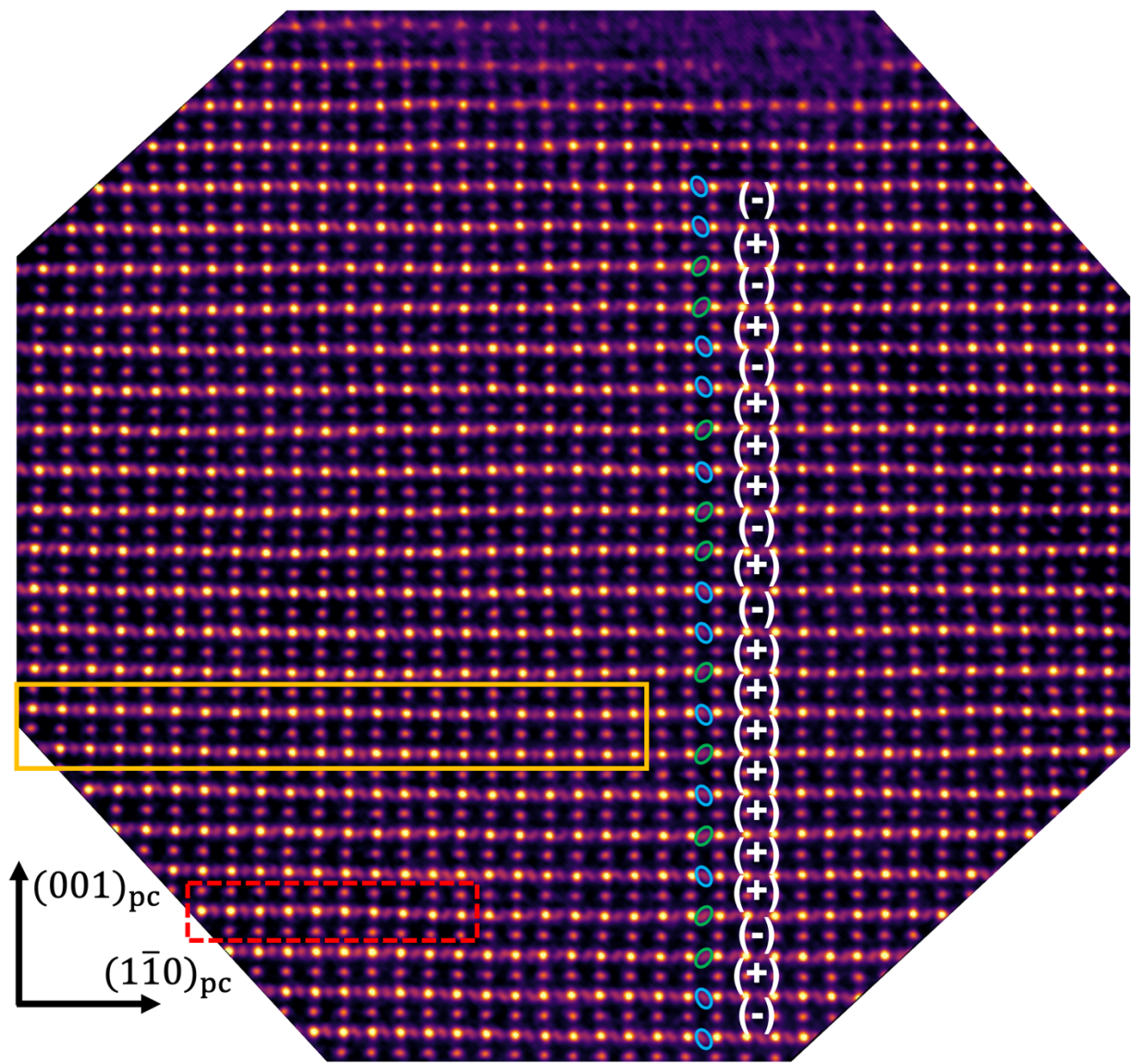

**Fig. S7** | The oxygen octahedral tilt pattern along the growth direction is mapped out over a larger field of view showing its incommensurate nature. In the region shown within the red dotted box, there is a switch from zig-zag in the same $(001)_{pc}$ plane, representative of a domain boundary for the in-plane oxygen octahedral rotations. In addition, an orange box marks a region where there does not appear to be a clear zig/zag pattern, likely indicative of a local deviation from the out-of-phase tilt pattern along the in-plane directions.

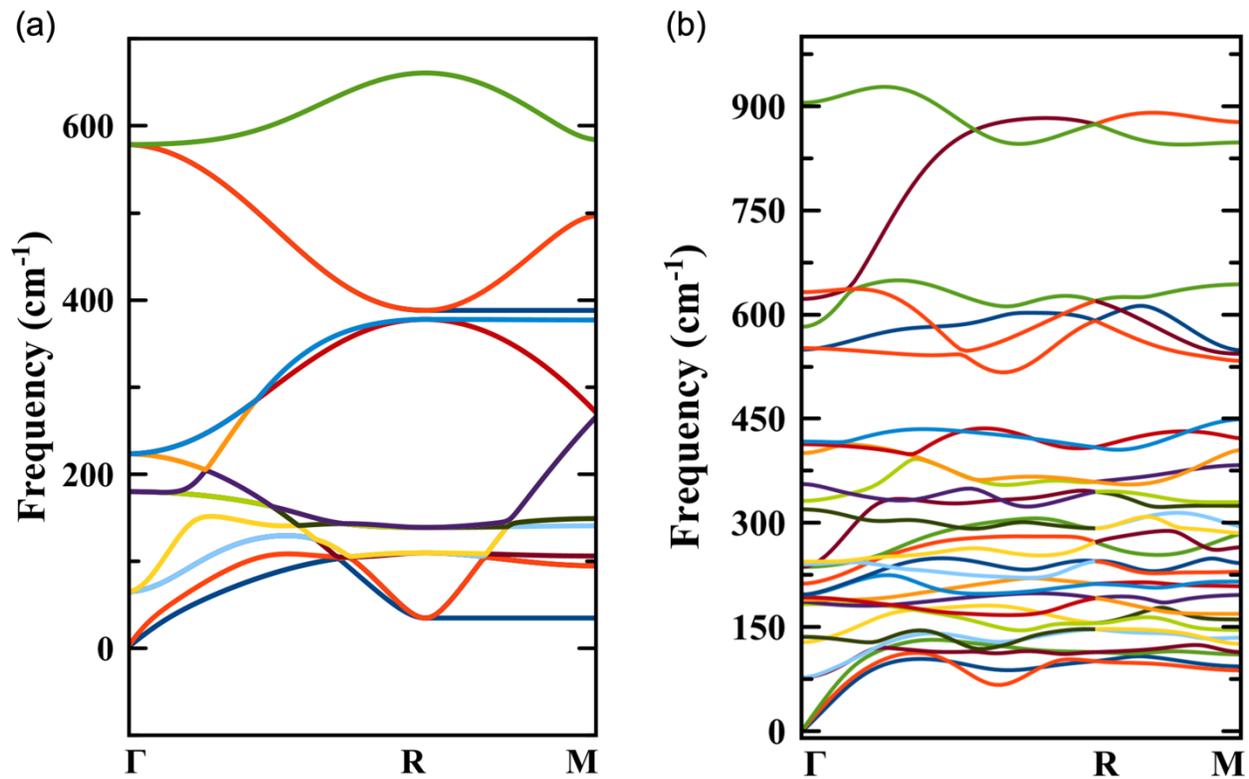

**Fig. S8** | Phonon band structure for the (a) cubic phase of sodium niobate calculated for 300 K using DFT VASP and TDEP[2] program and (b) *Cc* phase calculated for 0 K.

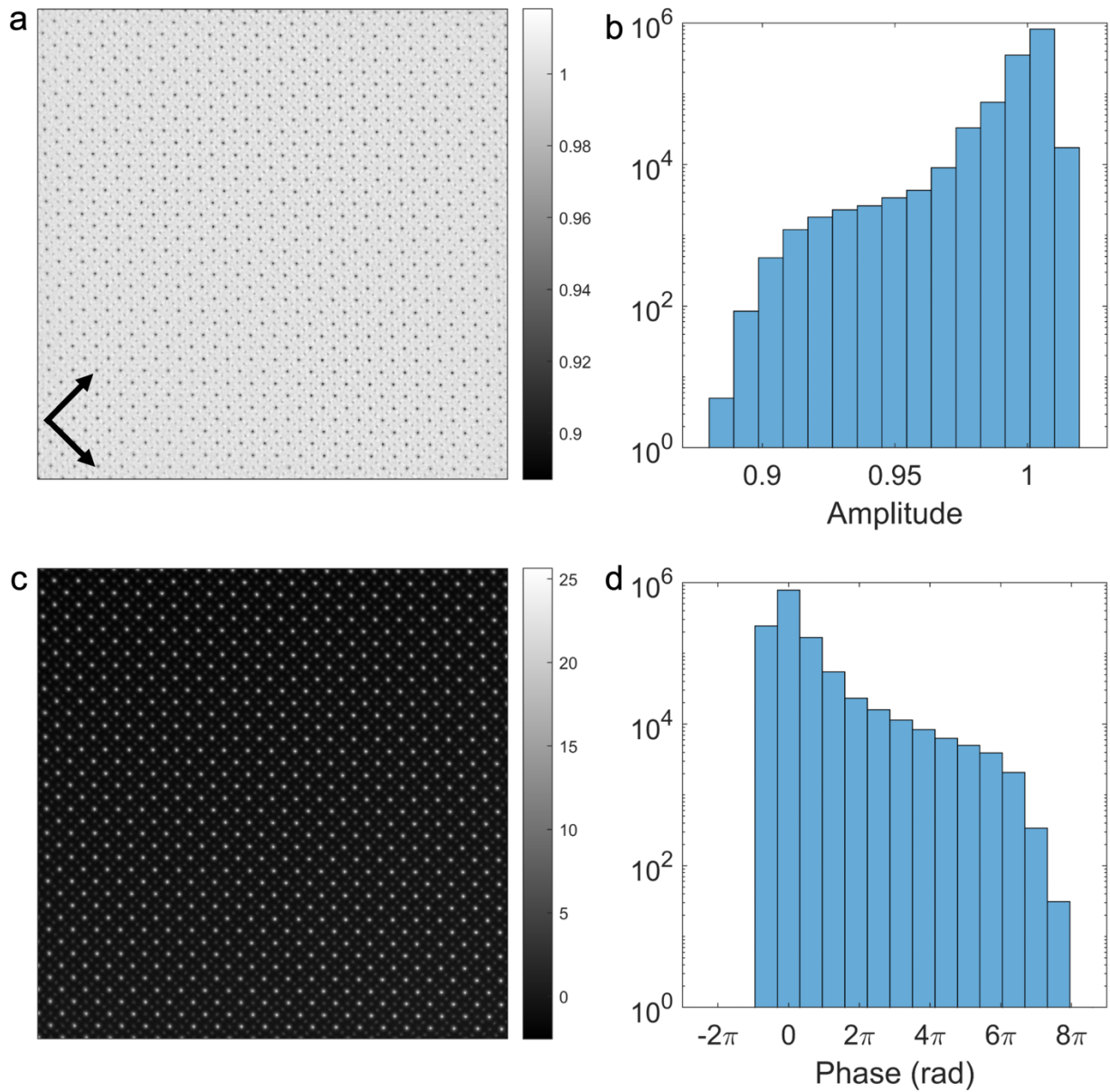

**Fig. S9** | Amplitude (a) and Phase (c) of the reconstructed object from an example multislice ptychography reconstruction of the sodium niobate sample. The pseudo-cubic directions of sodium niobate are labelled with black arrows. The histograms of the amplitude (b) and phase (d) are also shown. The amplitude image is the average of the amplitude from all slices. The phase image is the summation of the phase from all slices. Since each slice in the reconstruction is assumed to correspond to a phase object, the object amplitude for each slice should remain close to unity. Deviations from unity primarily correspond to scattering beyond the outer collection angle of the detector.

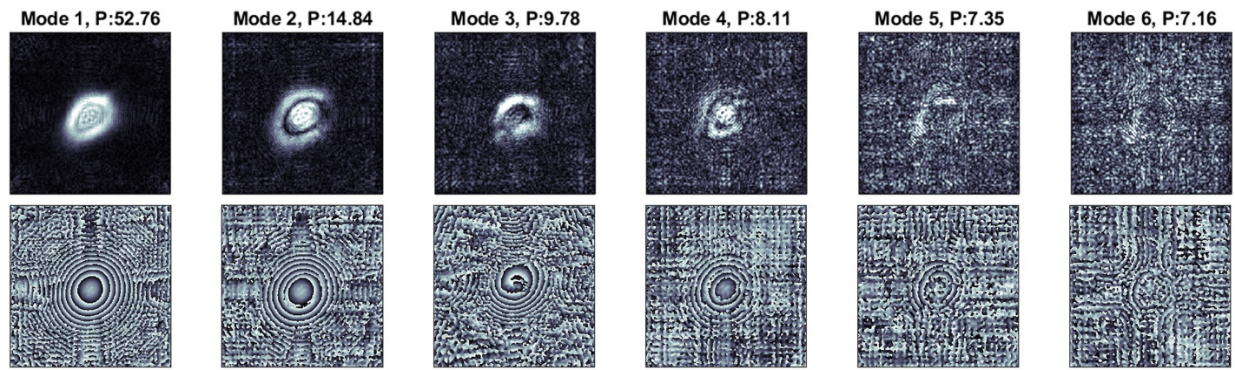

**Fig. S10** | Amplitude (top panel) and phase (bottom panel) for the six different probe modes used in the reconstruction for the dataset shown in Fig. S9. The index of each probe mode and its fractional intensity contribution (P) to the overall probe is also labelled. Note the residual astigmatism in the reconstructed probe, demonstrating the ability of the ptychographic algorithm to reconstruct the object and probe, even in the presence of residual aberrations.

<u>**Supplementary Text**</u>

1. **The different phases of sodium niobate**

Despite being studied for more than half a century, the crystal structure, origins of the ferroelectric/antiferroelectric distortions and phase transitions in $NaNbO_3$ are not completely understood. Early X-ray diffraction work identified seven different phases in bulk $NaNbO_3$ as a function of temperature, primarily characterized by differences in the oxygen octahedral tilt pattern and Nb displacements from the octahedron centers[3–5]. The phase transition sequence along with the reciprocal space points associated with the softened phonon modes as reported by Mishra et al.,[6] is given below:

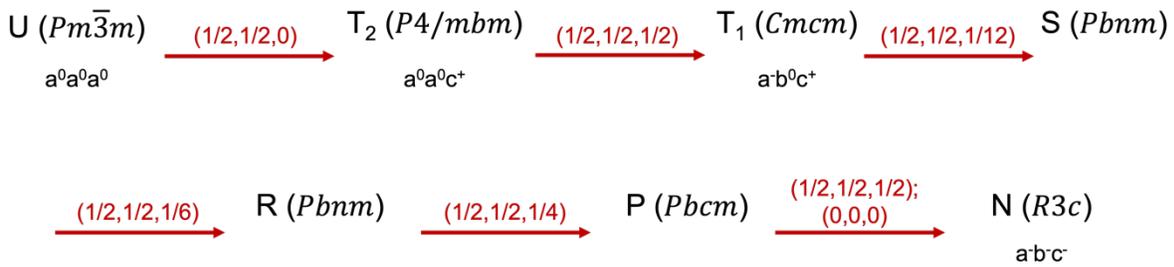

All these phonon modes (except the zone-center mode) involve rotations of the oxygen octahedra, which serve as the primary order parameter governing the phase transitions. Among these, the $M_3^+$ (1/2,1/2,0) / $R_4^+$ (1/2,1/2,1/2) modes cause an in-phase / anti-phase tilting of the octahedra and the resulting tilt pattern is denoted in the Glazer notation. The other soft phonon modes lie on the T line connecting the M (1/2,1/2,0) and R (1/2,1/2,1/2) points of the Brillouin zone and lead to complex tilt patterns along the c-axis. Such transitions are intimately related to the flat nature of the phonon bands in the M-R direction[7] which is responsible for the incommensurate tilt patterns reported in this study.

Although these seven phases continue to be consistently referenced in literature, several subsequent studies using X-ray, electron and neutron diffraction have contested the phase transition sequence, questioned the symmetry assigned to certain phases, and proposed new phases, including incommensurate ones. For example, the room temperature phase of NaNbO$_3$ is still subject to dispute with claims of *Pbcm*, *Pmc2$_1$*, *Pm* and *Pca2$_1$* space group symmetries[3,8–11]. Moreover, with several reports of multiple co-existing phases in NaNbO$_3$ at room temperature[3,11–15] and the small domain sizes observed in PFM, a local probe like that used in this study is better suited to identify and answer the questions of local site symmetries in phase-pure regions, as opposed to the spatially averaged measurements from X-ray or neutron diffraction.

Despite disputes over the ground state symmetry, NaNbO$_3$ is widely considered to be an antiferroelectric at room temperature and can be transmuted to a ferroelectric in thin film form by applying biaxial compressive or tensile strain[16,17]. We highlight a previous neutron powder diffraction study that identified oxygen displacements as the primary reason for the spontaneous polarization in the monoclinic *Pm* phase of NaNbO$_3$[10], similar to our observations for the new *Pc* and *Cm* phases of NaNbO$_3$. Free-standing membranes of sodium niobate have also been demonstrated to have a thickness dependent transition between ferroelectric (< 40 nm) to antiferroelectric (> 268 nm) ground states[12]. The size-dependent and strain induced tunability of the polarity and device-relevant magnitude of the spontaneous polarization in the ferroelectric state makes sodium niobate a very promising alternative for lead-based ferroelectrics and anti-ferroelectrics[18–20].

## 2. Distortions in terms of symmetry modes of the undistorted perovskite

Starting with the parent undistorted perovskite structure with $Pm\bar{3}m$ symmetry, we calculate the distortions that lead to the final DFT-calculated structures in terms of symmetry modes of the parent structure. This decomposition is performed using the ISODISTORT tool[21,22].

For the ***Pc* structure** (with an in-plane lattice constant of ~3.95 Å), the major contributing displacement modes and the corresponding parent-cell-normalized amplitude (normalized to the displacement in one pseudo-cubic unit cell) in Å for the different atoms are:

1) $\Gamma_4^-$ is the zone center mode that breaks inversion symmetry leading to the polarization. The mode comprises large displacements of the oxygen atoms and negligible displacements of the cations, consistent with our experimental measurements.

|  | x | y | z |
|---|---|---|---|
| Na (T1u) | 0 | 0 | 0 |
| Nb (T1u) | 0.0014 | 0.0013 | 0 |
| O (A2u) | -0.1151 | -0.1151 | 0 |
| O (Eu) | -0.1834 | -0.1835 | 0 |

2) $R_4^+$ is the mode that drives the anti-phase tilting of the oxygen octahedra along the pseudo-cubic c-axis.

|  | $a^-b^0c^0$ | $a^0b^-c^0$ | $a^0b^0c^-$ |
|---|---|---|---|
| O (Eu) | 0 | 0 | 0.2953 |

3) $M_3^+$ is the mode that drives the in-phase tilting of the oxygen octahedra along the pseudo-cubic a- and b-axes.

|        | $a^+b^0c^0$ | $a^0b^+c^0$ | $a^0b^0c^+$ |
|--------|-------------|-------------|-------------|
| O (Eu) | 0.3760      | 0.3762      | 0           |

4) $M_5^-$ is the mode that generates the anti-polar displacements of the cations. The anti-polar displacements are mainly along the two in-plane pseudo-cubic directions for Na and along the out-of-plane pseudo-cubic direction for Nb,

|          | a       | b       | c       | d | e | f       |
|----------|---------|---------|---------|---|---|---------|
| Na (T1u) | -0.0336 | -0.0336 | -0.0523 | 0 | 0 | -0.0524 |
| Nb (T1u) | -0.0029 | -0.0029 | -0.0229 | 0 | 0 | -0.0229 |
| O (Eu)   | 0.0086  | 0.0086  | -0.0004 | 0 | 0 | -0.0004 |

The above four symmetry modes account for 99.2% of the total distortions from the undistorted perovskite structure (excluding strain modes).

Similarly, for the ***Cm* structure** (with an in-plane pseudo-cubic lattice constant of ~3.945 Å), the major contributing displacement modes and the corresponding parent-cell-normalized amplitude (in Å) for the different atoms are:

1) $\Gamma_4^-$ mode

|          | x | y | z |
|----------|---|---|---|
| Na (T1u) | 0 | 0 | 0 |
| Nb (T1u) | -0.0253 | -0.0152 | 0 |
| O (A2u) | -0.1478 | -0.1327 | 0 |
| O (Eu) | -0.2365 | -0.2129 | 0 |

2) $M_3^+$ mode

|  | $a^+b^0c^0$ | $a^0b^+c^0$ | $a^0b^0c^+$ |
|---|---|---|---|
| O (Eu) | 0.3497 | 0.3535 | 0.3357 |

3) $M_5^-$ mode

|  | a | b | c | d | e | f |
|---|---|---|---|---|---|---|
| Na (T1u) | 0.0314 | -0.0909 | -0.0379 | 0 | 0 | -0.0859 |
| Nb (T1u) | 0.0181 | -0.0245 | -0.0202 | 0 | 0 | -0.0250 |
| O (Eu) | -0.0050 | 0.0185 | -0.0034 | 0 | 0 | 0.0088 |

The above three symmetry modes account for 99.6% of the total distortions from the undistorted perovskite structure (excluding strain modes).

## 3. Calculation of polarization for the *Pc* and *Cm* structures

The Born effective charge (BEC) for each atom is calculated by displacing it along the x, y, and z directions by 0.01 Å and computing the dipole moment that arises from this displacement. The polarization is then calculated using the computed BECs and the displacements of different ions with respect to a reference centrosymmetric structure.

For the *Pc* structure with an in-plane pseudo-cubic lattice constant of 3.945 Å, the BECs are:

|      | x     | y     | z     |
|------|-------|-------|-------|
| Na   | 1.12  | 1.12  | 1.10  |
| Nb   | 9.21  | 9.21  | 9.51  |
| O1   | -1.66 | -1.66 | -1.65 |
| O2   | -7.04 | -7.04 | -7.35 |

The calculated value for the total polarization is 43.82 $\mu C/cm^2$ ($P_x$ = 30.98 $\mu C/cm^2$, $P_y$ = 30.98 $\mu C/cm^2$, $P_z$ = 0 $\mu C/cm^2$)

For the *Cm* structure with an in-plane pseudo-cubic lattice constant of 3.945 Å, the BECs are:

|      | x     | y     | z     |
|------|-------|-------|-------|
| Na   | 1.12  | 1.12  | 1.10  |
| Nb   | 9.23  | 9.23  | 9.48  |
| O1   | -1.67 | -1.67 | -1.65 |
| O2   | -7.05 | -7.05 | -7.32 |

The calculated value for the total polarization is 46.31 $\mu C/cm^2$ ($P_x$ = 33.52 $\mu C/cm^2$, $P_y$ = 31.95 $\mu C/cm^2$, $P_z$ = 0 $\mu C/cm^2$)